\documentclass[12pt, a4paper]{article}

 \usepackage[font=small,format=plain,labelfont=bf,up,textfont=normal,up,justification=justified,singlelinecheck=false]{caption}
\usepackage{subcaption}
\usepackage{mathrsfs}
\usepackage[a4paper, left=2cm,right=2cm]{geometry}
\usepackage[colorlinks=true,linkcolor=black,citecolor=teal,urlcolor=MidnightBlue,filecolor=black]{hyperref}
\usepackage{amsfonts}
\usepackage{amsmath,amssymb}
\usepackage{setspace}
\usepackage{slashed}
\usepackage{braket}
\usepackage{color}
\usepackage{amsmath}
\usepackage[dvipsnames]{xcolor}
\definecolor{SchoolColor}{rgb}{0.6471, 0.1098, 0.1882} 
\usepackage{subfloat}

\usepackage{tensor}
\usepackage{cite}
\usepackage{tikz,tikz-cd,tkz-euclide,tikz-feynman}
\usetikzlibrary{calc}
\usetikzlibrary{patterns}
\usetikzlibrary{arrows.meta}
\usetikzlibrary{intersections}
\tikzfeynmanset{warn luatex=false}
\usetikzlibrary{arrows.meta}
\usetikzlibrary{positioning}
\usetikzlibrary{arrows.meta}

\usepackage{graphicx}
\usepackage{bm} 

\usepackage{fontspec}
\usepackage{tensor}
\usepackage{cite}
\usepackage{graphicx}
\graphicspath{{figure/}}
\usepackage{array}
\usepackage{booktabs}

\usepackage{multirow}

\usepackage{dcolumn}
\usepackage{bm}

\usepackage{verbatim}

\usepackage{textcomp} 
\usepackage{graphicx} 

\numberwithin{equation}{section}

\usepackage{pgfplots}

\newcommand{\bea}{\begin{eqnarray}}
\newcommand{\eea}{\end{eqnarray}}
\newcommand{\be}{\begin{equation}}
\newcommand{\ee}{\end{equation}}
\def\nn{\nonumber}

\def\d{\mathrm{d}}

\def\0{{(0)}}\def\1{{(1)}}\def\2{{(2)}}\def\3{{(3)}}\def\4{{(4)}}
\def\mn{{\mu\nu}}
 
\newcommand{\Rn}[1]{{\rm\uppercase\expandafter{\romannumeral#1}}}
 
\def\nn{\nonumber}

\def\qaq{\quad\text{and}\quad}

\def\qwq{\quad\text{with}\quad}

\def\c.c.{\mathrm{c.c.}}

\def\g{\gamma}

\def\th{\theta}

\def\m{\mu}\def\n{\nu}

\def\ci{\mathcal{I}}

\begin{document}

\begin{titlepage}

\begin{flushright}\vspace{-3cm}
{\small
\today }\end{flushright}
\vspace{0.5cm}
\begin{center}
	{{ \LARGE{\bf{
Electromagnetic helicity flux density for radiative systems  }}}}\vspace{5mm}

	\centerline{\large{\bf 
	Zhen-Yu Heng\footnote{zhenyuheng@hust.edu.cn},
Jiang Long\footnote{longjiang@hust.edu.cn},
Run-Ze Yu\footnote{yurunze01@hust.edu.cn}\textsuperscript{}\textsuperscript{*},
\& Xin-Hao Zhou\footnote{zhouxinhao01@hust.edu.cn}
\renewcommand{\thefootnote}{\fnsymbol{footnote}}{\footnotetext[1]{Author to whom any correspondence should be addressed.}}\renewcommand{\thefootnote}{\arabic{footnote}}
	}}
	\vspace{2mm}
	\normalsize
	\bigskip\medskip

	\textit{School of Physics, Huazhong University of Science and Technology, \\ Luoyu Road 1037, Wuhan, Hubei 430074, China
	}
	
	\vspace{25mm}
	
	\begin{abstract}
		\noindent
		{We show that the helicity flux density is distinguished from magnetic helicity by analysing Hopf solitons. The electromagnetic (EM) helicity flux and the magnetic helicity are Chern-Simons terms at different hypersurfaces. We find the helicity flux density for a point charge moving with an acceleration, extending the Li\'enard-Wiechert angular distribution of radiant power. We also derive the multipole expansion of the helicity flux density, generalizing the Larmor's formula for the radiant power. These formulae have been applied to discuss the helicity flux density in  several toy models such as circular and helical motion as well as soft bremsstrahlung. We also comment on the potential applications  of the EM helicity flux density to pulsar systems.}\end{abstract}
	

\end{center}

\end{titlepage}

\tableofcontents

\section{Introduction}\label{cpsf}
It is well known that waves transfer energy, momentum, and angular momentum as they propagate.
These physical processes typically originate from electromagnetic radiation, as extensively discussed in \cite{jackson1999classical}. For gravitational radiation, the corresponding energy loss  was observed indirectly in a pulsar binary system, PSR B1913+16 by Hulse and Taylor  half a century ago \cite{Hulse:1974eb,1982ApJ...253..908T}. 
Interestingly, according to the representation theory of the  Poincar\'e group
\cite{Wigner:1939cj,1948ZPhy..124..665W,Bargmann:1948ck,weinberg_1995}, mass and spin are two important quantum numbers at the microscopic level. In particular, for massless representations, the spin quantum number is replaced by helicity, an integer or half-integer defined as the projection of the spin onto the direction of the  momentum. Although there is much discussion on the radiation of energy and angular momentum, the research on the radiation of helicity remains scarce.

Recently, the gravitational helicity flux density in a two-body system has been discussed in the post-Newtonian framework \cite{Long:2024yvj}. The main conclusion is that the angular distribution of the helicity flux is non-trivial, despite the fact that the total radiative helicity is zero after integrating over the celestial sphere. Unfortunately, from a practical point of view, the signal generated by a typical binary black hole system is diluted due to the large distance to the observer on Earth. In the foreseeable future, it is  unlikely to be detected via this non-linear spin precession of a free-falling  gyroscope \cite{Seraj:2022qyt, Faye:2024utu} in the solar system.

\begin{figure}[h!]
    \centering
    \includegraphics[width=0.6\textwidth]{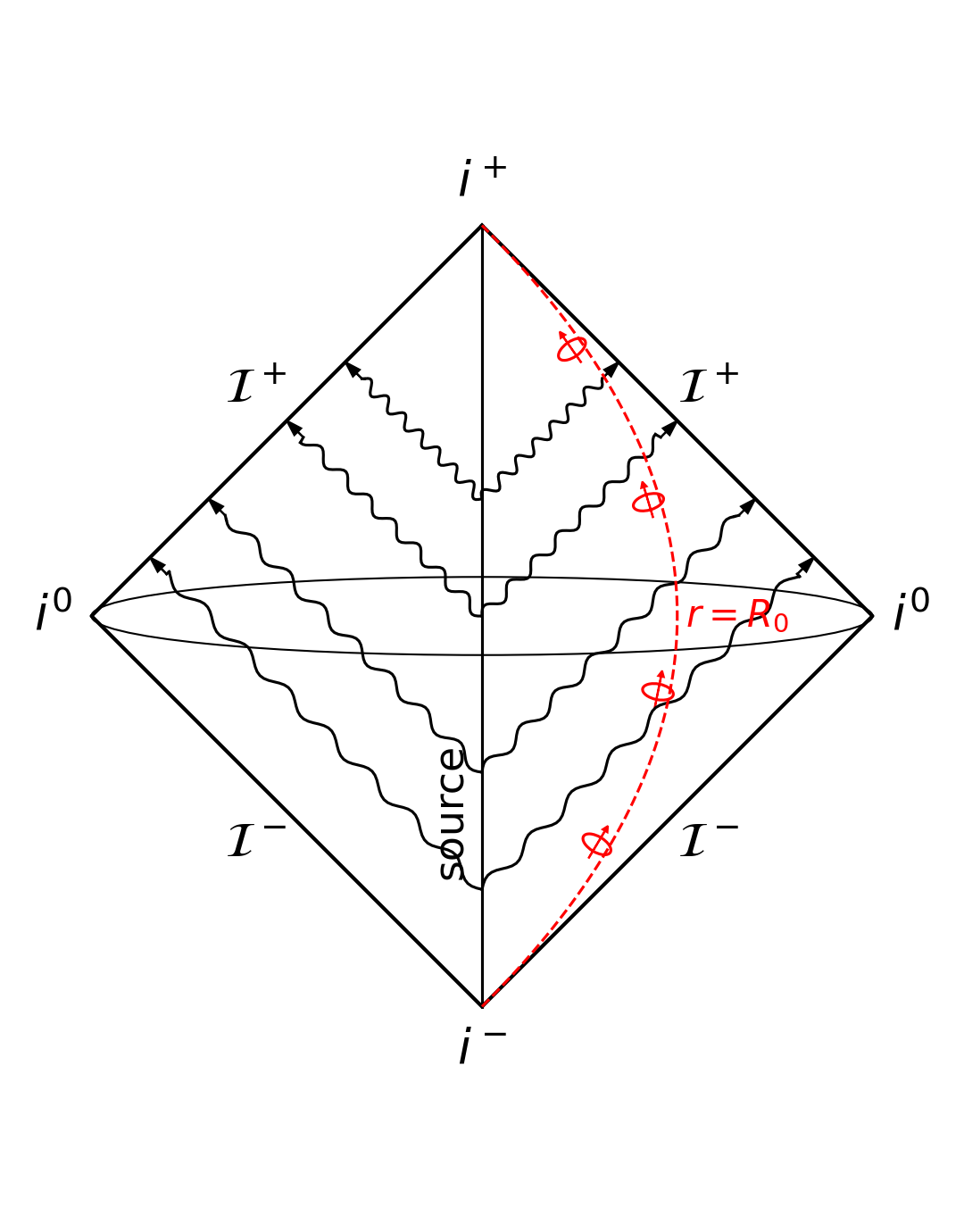}
    \caption{ This figure schematically illustrates a method for measuring the EM radiative helicity flux density within the Penrose diagram of Minkowski spacetime. An EM wave, generated by a source, propagates through the bulk  and interacts with a detector (e.g., a magnetic dipole) positioned near  $\mathcal I^+$, inducing a precession of the dipole. In the diagram, the magnetic dipole is depicted as a circular loop of electric current, with an arrow indicating its direction according to  the right-hand rule. The dashed line represents the worldline of the detector at a constant radius $r=R_0$. A similar setup, employing a free-falling gyroscope as the detector to measure the gravitational helicity flux density, can be found in \cite{Seraj:2022qyt}.}
    \label{detect}
\end{figure}

While the detection of gravitational helicity flux density is challenging, we recognize that electromagnetic radiation offers another, critically important, avenue of study in astrophysics. Pulsars, predicted in the early 1930s \cite{Landau:1932uwv, 1934PNAS...20..259B}, and detected in 1967 \cite{Hewish:1968bj},  are magnetized neutron stars that emit periodic short pulses of radio radiation. Recently, the EM helicity flux density has been constructed in \cite{Liu:2023qtr} and its physical effect has been explored in \cite{Oblak:2023axy,Maleknejad:2023nyh}. {As conceptually shown in Fig.\ref{detect}, this can be understood as the ``chiral memory effect'', where the radiation from the source carrying helicity interacts with a test object such as a magnetic dipole \cite{Oblak:2023axy}. This  provides a physical mechanism for measuring the helicity flux density.} 

The topological nature of the radiative helicity has been investigated by \cite{Liu:2024rvz} and its connection to chiral anomaly has been mentioned in \cite{Guo:2024qzv}. We should mention that all the results are achieved in Bondi's coordinate systems \cite{Bondi:1962px} and are closely related to the geometry and topology of future null infinity $\mathcal I^+$, a  well-known Carrollian manifold \cite{Une,Gupta1966OnAA,Duval_2014a,Duval_2014b,Duval:2014uoa}.  Carrollian symmetries have been applied to flat holography  \cite{Donnay:2022aba,Bagchi:2022emh} successfully and 
play a central role in the study of  ultra-relativistic hydrodynamics  \cite{Ciambelli:2018xat,Petkou:2022bmz,Donnay:2019jiz,Freidel:2022vjq,Redondo-Yuste:2022czg}. These topics have also been uplifted to gravitational theories and particle dynamics. Gravity in the ultra-relativistic  limit, called Carroll gravity,  has ``black hole solutions'' \cite{Hansen:2021fxi,Ecker:2023uwm}. 
The helicity flux operator deforms the Carrollian diffeomorphism \cite{Ciambelli:2018xat} into  the so-called intertwined Carrollian diffeomorphism \cite{Liu:2023qtr,Liu:2023gwa} and characterizes the angular distribution of the difference  in the numbers of massless particles with left and right hand helicities. It corresponds to the duality invariance of the theory \cite{Dirac:1931kp,Deser:1976iy,Henneaux:2004jw,Godazgar:2018qpq} in the bulk, and generates the superduality transformation of the fundamental field at the null boundary. These remind us of a similar concept, which is known as magnetic helicity \cite{elsasser1956hydromagnetic,woltjer1958theorem} in plasma, fluid and astrophysics. However, we could show that they are closely related but not equivalent using EM Hopfion solutions \cite{arrayas2017knots}. Actually, the helicity flux density provides much more detailed  information  about the system, rather than just topological aspects. Thus, the helicity flux density is an interesting new physical observable that deserves to explore its properties in real radiative systems.

In this work, we will  systematically study the EM helicity flux density radiated to $\mathcal I^+$ from general bulk sources. Similar to the  radiative gravitational helicity, a total radiative EM helicity is always zero for a point particle. However, we still find interesting angle-distributions  of the helicity flux density in several toy models. For  readers  interested in astrophysics, we also discuss the possible applications of the  formulae to the pulsar systems at the end of this paper.

The layout of the paper is as follows. We will review the necessary formalism used in this work in  section \ref{formalism}. This is shown by carefully introducing the coordinate systems and defining the helicity flux density operators. We will also discuss the main differences and connections among helicity flux density, magnetic helicity and optical helicity in the same section. In section \ref{ems}, we investigate the general aspects of EM helicity flux density and apply the formulae to four toy models. In the following section, we study the multipole expansion of the helicity flux for completeness. We will close this paper by exploring the helicity flux density in pulsar systems. Lengthy computations are presented in three appendices.

\section{Review of the formalism}\label{formalism}
\subsection{Coordinate systems}
We now introduce the conventions used in this paper. In flat spacetime $\mathbb R^{1,3}$, it is convenient to describe the future null infinity ($\mathcal{I}^+$) in the retarded coordinates $x^\alpha=(u,r,\theta^A)$ with $\theta^A=(\th,\phi)$ the spherical coordinates for the unit sphere. The metric  of $\mathbb R^{1,3}$ reads 
\begin{align}
    d s^2=-d u^2-2d ud r+r^2(d\th^2+\sin^2\th d\phi^2).\label{metric}
\end{align}
The retarded coordinates are converted to the Cartesian coordinates $x^\mu=(t,x^i)$ through
\begin{align}
    x^\mu=u\bar m^\m+r n^\mu \qwq \bar m^\m=(1,0)=\frac{1}{2}(n^\mu-\bar n^\mu)
\end{align} 
where we have already defined two null vectors $n^\mu$ and $\bar n^\mu$ 
\begin{align}
    n^\mu=(1,n^i),\qquad \bar n^\mu=(-1,n^i) \label{null}
\end{align}
with $n^i$ the normal vector of the unit sphere 
\begin{align}
    n^i=(\sin\theta\cos\phi,\sin\theta\sin\phi,\cos\theta).
\end{align}

We can further introduce 
\begin{align}
    m^\m=\frac{1}{2}(n^\mu+\bar n^\mu)=(0,n^i) \qaq Y^A_\m=-\nabla^An_\mu\label{strictckv}
\end{align}
where the covariant derivative $\nabla^A$ is adapted to the metric on the null boundary
\begin{align}
    d s^2_{\ci^+}=d s^2_{S^2}=d\th^2+\sin^2\th d\phi^2\equiv \g_{AB}d \theta^A d \theta^B.
\end{align}
The six conformal Killing vectors (CKVs) on the sphere are collected as $Y^A_\mn=Y^A_\m n_\n-Y^A_\n n_\m$ whose properties can be found in the appendix of \cite{Liu:2023gwa}. The integral measures are denoted by
\begin{align}
    \int d ud\Omega\equiv\int_{-\infty}^{\infty}d u\int_0^{2\pi}d\phi\int_0^\pi\sin\th d\th
\end{align} on $\ci^+$ and  
\begin{align}
\int d\Omega\equiv \int_0^{2\pi}d\phi \int_0^\pi \sin\theta d\theta.
\end{align} on the unit sphere. The metric of a spacelike hypersurface $V$ with constant Cartesian time is denoted by
\begin{align} 
d s_V^2=\delta_{ij}d x^i d x^j=d x^2+d y^2+d z^2
\end{align} and the integral measure on this hypersurface is 
\begin{align} 
\int_V d^3\bm x.
\end{align} 
\subsection{Electromagnetic theory and its asymptotic expansion}
The electromagnetic vector potential is a four-component (co-)vector $a_\mu$ and the electric and magnetic fields are combined into a skew-symmetric  tensor 
\bea  
f_{\mu\nu}=\partial_\mu a_\nu-\partial_\nu a_\mu.\label{fmunu}
\eea  To be more precise, the electric field $e_i$ and magnetic field $b_i$ are  
\bea 
e_i=-f_{0i},\quad b_i=\frac{1}{2}\epsilon_{ijk}f^{jk},
\eea where the symbol $\epsilon_{ijk}$ denotes the Levi-Civita tensor in three dimensions with the convention $\epsilon_{123}=1$ in Cartesian coordinates. The expression \eqref{fmunu} can be written compactly as 
\be 
f=d a
\ee where $a=a_\mu dx^\mu$ is a one-form and $f=\frac{1}{2}f_{\mu\nu}dx^\mu\wedge dx^\nu$ is a two-form.  The action is 
\bea 
S=\int d^4x \sqrt{-g}[-\frac{1}{4}f_{\mu\nu}f^{\mu\nu}+j_\mu a^\mu], \label{action}
\eea where the last term involves a source $j_\mu$ coupled to the vector field $a_\mu$ and causes the EM radiation.  The gauge transformation of the vector field is 
\be 
\delta_\eta a_\mu=\partial_\mu\eta
\ee while the EM field $f_{\mu\nu}$ is invariant under the gauge transformation. The current is conserved 
\be 
\partial_\mu j^\mu=0
\ee to save the gauge invariance of the action. 
The Maxwell equations are derived from the variational principle
\bea 
\partial_\mu f^{\mu\nu}=-j^\nu \label{maxeq}
\eea and the Bianchi identity
\be
\partial_{[\mu}f_{\rho\sigma]}=0.
\ee In these equations, the partial derivative $\partial_\mu$ is adapted to the flat metric in Cartesian coordinates. In real systems, the current $j^\mu$ is located in a finite domain of spacetime and one may set it to zero near $\mathcal{I}^+$. However, it will affect the field near $\mathcal I^+$ via Green's function. For a radiative field, the fall-off behaviour near $\mathcal I^+$ is 
\be 
a_\mu(x)=\frac{A_\mu(u,\Omega)}{r}+\cdots
\ee which can be transformed to the components in retarded coordinates 
\be 
a_u(x)=\frac{A_u(u,\Omega)}{r}+\cdots,\quad a_r(x)=\frac{A_r(u,\Omega)}{r}+\cdots,\quad a_A(x)=A_A(u,\Omega)+\cdots
\ee where the transverse component $A_A(u,\Omega)$ is a projection of the spatial one
\be 
A_A(u,\Omega)=-Y^i_A A_i(u,\Omega). 
\ee Note that the projector $Y^i_A$ has been defined in  \eqref{strictckv}.  The asymptotic expansion of the electric and magnetic fields are
\be 
e_i(x)=\frac{E_i(u,\Omega)}{r}+\cdots,\quad b_i(x)=\frac{B_i(u,\Omega)}{r}+\cdots
\ee where the leading coefficients $E_i$ and $B_i$ are fixed by the time derivative of the transverse modes 
\be 
E_i(u,\Omega)=Y_i^A \dot A_A(u,\Omega),\quad B_i(u,\Omega)=-\tilde{Y}_i^A \dot A_A(u,\Omega).
\ee We have defined $\dot{A}_A\equiv\partial_uA_A$ and the Hodge dual of $Y_i^A$ on the unit sphere
\be 
\tilde{Y}_{i}^{\ A}=Y_{iC}\epsilon^{CA}
\ee where $\epsilon_{AB}$ is the two dimensional Levi-Civita tensor 
\begin{align}
    \epsilon_{AB}=\begin{pmatrix}
        0 & \sin\theta \\ -\sin\theta & 0
    \end{pmatrix}.
\end{align}
\subsection{Flux densities}
In \cite{Liu:2023qtr}, the authors defined three flux densities at $\mathcal I^+$
\begin{align}
    T(u,\Omega)&=\dot A_A \dot A^A,\\
    M_A(u,\Omega)&=\frac{1}{2}(\dot A^B\nabla^C A^D-A^B\nabla^C\dot A^D) P_{ABCD},\\
    O(u,\Omega)&=\epsilon_{AB}A^A\dot A^B
\end{align} where the rank-4 tensor $P_{ABCD}$ is
\bea 
P_{ABCD}=\gamma_{AB}\gamma_{CD}+\gamma_{AC}\gamma_{BD}-\gamma_{AD}\gamma_{BC}.
\eea The energy flux density $T(u,\Omega)$ measures the radiant power of the energy $E$ per unit solid angle 
\bea 
\frac{d E}{d u d\Omega}=T(u,\Omega)
\eea while the angular momentum flux density $M_A(u,\Omega)$ is connected to the radiant power of the angular momentum $J_{ij}$ per unit solid angle 
\be 
\frac{d J_{ij}}{d u d\Omega}=Y_{ij}^A M_A(u,\Omega)\equiv M_{ij}(u,\Omega).
\ee Similarly, the helicity flux density $O(u,\Omega)$ is the radiant power of the helicity $H$ per unit solid angle 
\be 
\frac{d H}{d u d\Omega}=O(u,\Omega).
\ee The helicity flux density is completely absent in the classical textbook \cite{jackson1999classical}, we will investigate its various aspects in this work. For later convenience, we will define 
\begin{align}
\frac{d H}{d u}&=\int d\Omega \frac{d H}{d u \d\Omega},\\
\frac{d H}{d\Omega}&=\int d u \frac{d H}{d u \d\Omega},\\
H&=\int d u d\Omega  \frac{d H}{d u d\Omega}
\end{align} and call them helicity flux, helicity density and radiative helicity respectively. On the other hand, the helicity flux density always refers to $O$, which is also $\frac{d H}{d u d\Omega}$. For a $T$-periodic system, the average helicity flux density is denoted as 
\be 
\langle O(u,\Omega)\rangle=\langle \frac{dH}{du d\Omega}\rangle=\frac{1}{T}\int_0^T du \frac{dH}{du d\Omega}.
\ee 
\subsection{Comparable concepts in literature}
In this subsection, we will clarify the connections and differences between helicity flux density and similar concepts in literature.
\subsubsection{Magnetic helicity}
\begin{figure}[h!]
\centering
\begin{subfigure}[t]{0.48\textwidth}
\centering
\includegraphics[width=\linewidth]{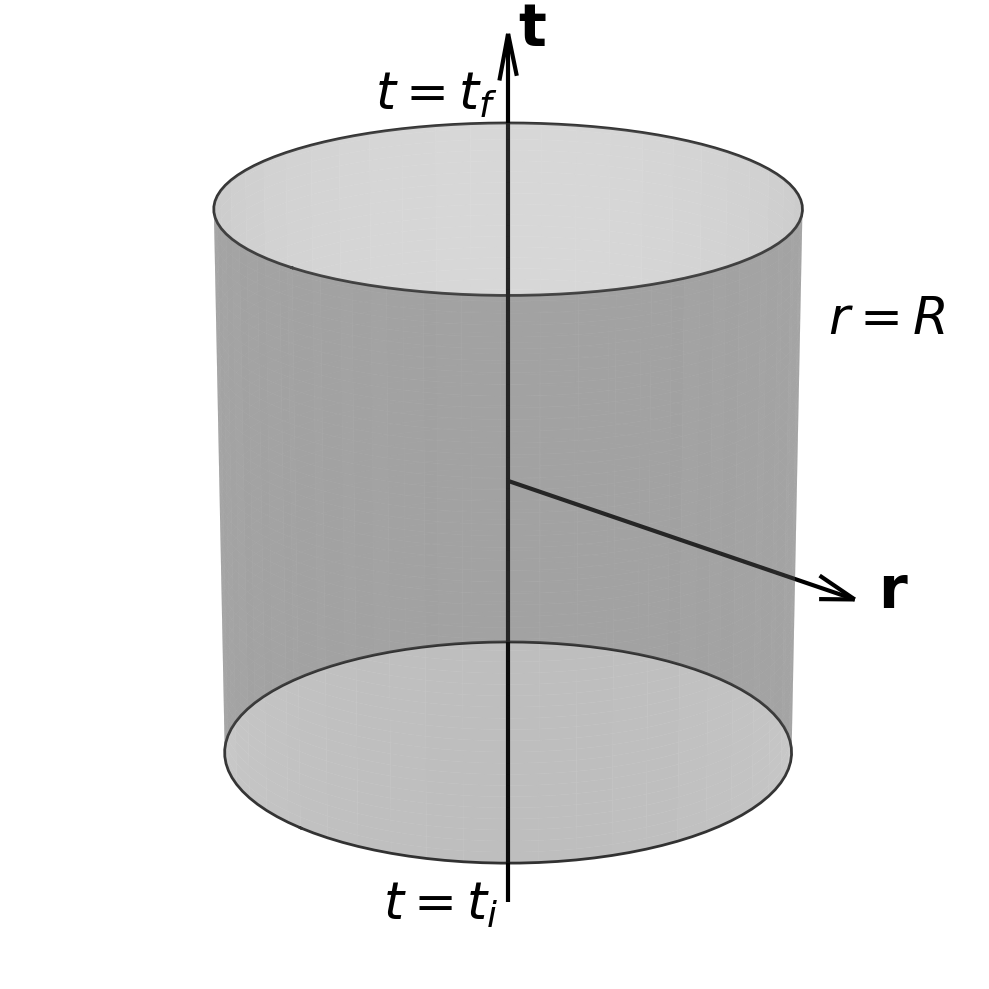}
\caption{Cylinder spacetime is bounded by two time slices, $t=t_f$ and $t=t_i$, as well as a timelike surface $r=R$. In the limit  $R\to+\infty,\,t_f\to+\infty,\,t_i\to-\infty$, this cylinder spacetime extends to cover the entire spacetime.}
\label{fig:rectangle_region}
\end{subfigure}
\hfill
\begin{subfigure}[t]{0.49\textwidth}
\centering
\includegraphics[width=\linewidth]{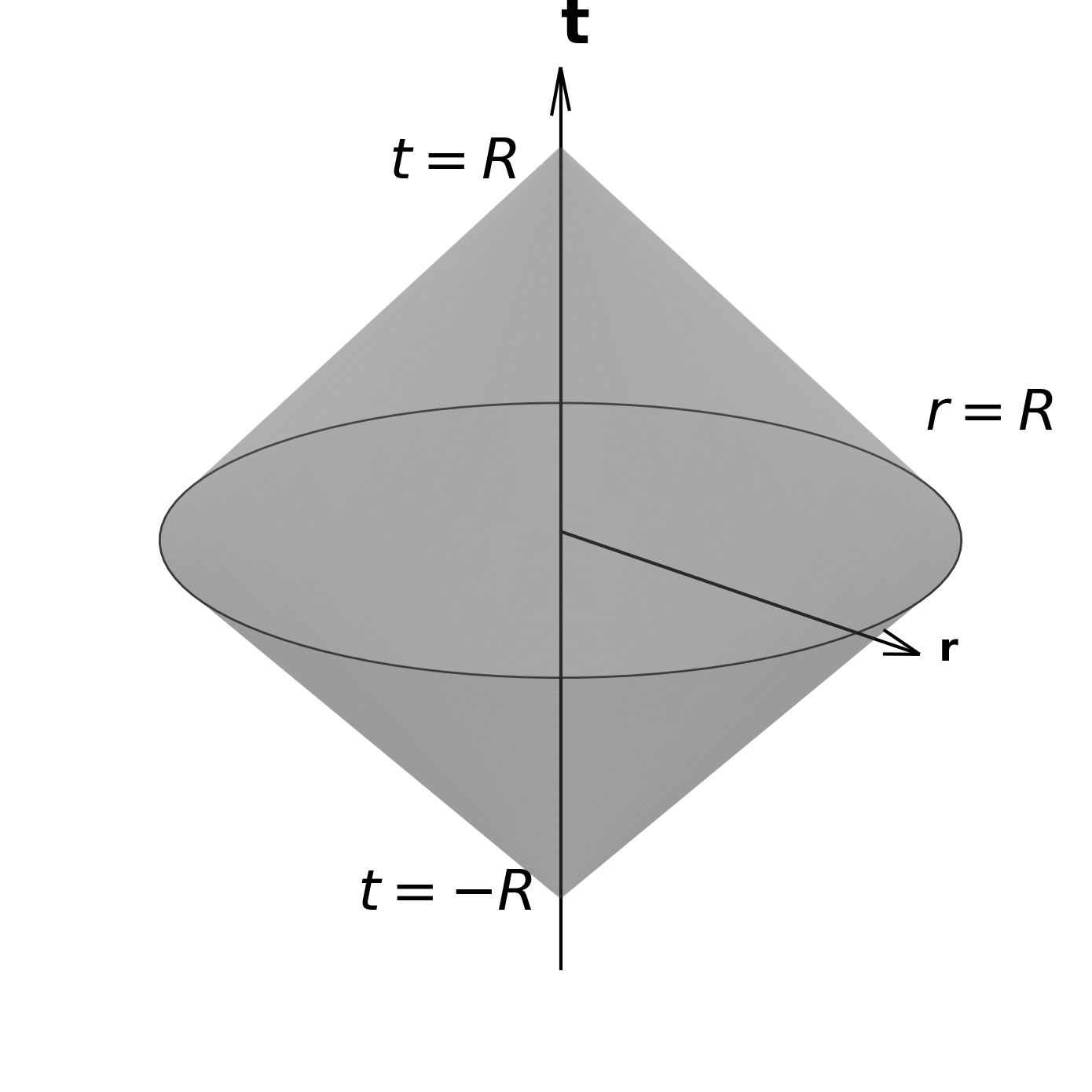}
\caption{Causal diamond with radius $R$. In the limit  $R\to+\infty$, the causal diamond extends to cover the entire spacetime.}
\label{fig:diamond1_coords}
\end{subfigure}
\caption{Two different approaches to cover the entire Minkowski spacetime}
\label{fig:two_regions_comparison}
\end{figure}

In astrophysics, it has been known for a long time that one can define magnetic helicity \cite{elsasser1956hydromagnetic,woltjer1958theorem} which describes dynamo processes and characterizes the rotation and twist of the vector field, providing a measure of
the linking number of the magnetic field lines. The magnetic helicity $\mathcal H_m$ is evaluated on a constant time slice $V$
\be 
\mathcal H_m=\int_V d^3\bm x\ \bm a\cdot\bm b
\ee where $\bm a$ represents the spatial component of the vector potential and $\bm b$ is the magnetic field. We also define magnetic helicity density as 
\be 
O_{m}(\bm x)=\bm a \cdot \bm b.\label{magdensity}
\ee As has been discussed in \cite{Long:2024yvj}, the radiative helicity $H$ defined in this paper is evaluated at $\mathcal I^+$
\be
H=\int du d\Omega \frac{dH}{du d\Omega}=\int du d\Omega\ \epsilon_{AB}A^A \dot A^B.
\ee There is no guarantee that they are the same. Actually, one can show that they are not identical in general. In \cite{ranada1989topological}, a static soliton in vacuum  Maxwell theory has been found using Hopf mapping \footnote{One of the early applications of the Hopf mapping in electromagnetism can be found in  \cite{Trautman:1977im}. The solitons  in magnetohydrodynamics (MHD) using Hopf mapping can be found in \cite{2004physics...9093K}. }
\begin{align}
\bm a(\bm x )
  &=\frac{4}{(1+r^{2})^{2}}
    \Bigl[(xz-y)\,\bm e_x
          +(x+yz)\,\bm e_y
          +\tfrac12\!\bigl(1-x^{2}-y^{2}+z^{2}\bigr)\,\bm e_z\Bigr].
          \end{align} As a consequence, the magnetic field is 
          \begin{align}
\bm b(\bm x)
  &=\nabla\times \bm a
  =\frac{16}{(1+r^{2})^{3}}
    \Bigl[(xz-y)\,\bm e_x
          +(x+yz)\,\bm e_y
          +\tfrac12\!\bigl(1-x^{2}-y^{2}+z^{2}\bigr)\,\bm e_z\Bigr].
\end{align}
It is straightforward to show that the magnetic helicity is a non-vanishing constant
\be 
\mathcal H_m=\int_V d^3\bm x\ \bm a\cdot\bm b=4\pi^2.
\ee On the other hand, the radiative helicity $H$ is vanishing  since there is no radiation   in this example. To find a non-vanishing helicity $H$, the solution must be dynamical, in contrast to the static case. In other words, we find 
\be 
\dot A_A=0\quad\Rightarrow\quad O=\epsilon_{AB}A^A\dot A^B=0\quad\Rightarrow\quad H=\int du d\Omega O(u,\Omega)=0.
\ee We conclude that the magnetic helicity $\mathcal H_m$ is not equal to the radiative helicity $H$
\be 
\mathcal H_m\not=H.
\ee 
However, one should admit that the two concepts are quite similar. The magnetic helicity can be shown as the Chern-Simons term on the spatial hypersurface $V$
\be 
\mathcal H_m=\int_V a\wedge da
\ee while the helicity $H$ is the Chern-Simons term at future null infinity \cite{Liu:2024rvz}
\be 
H=\int_{\mathcal I^+}a\wedge da.
\ee 
This provides the connection between the two concepts. We denote the Chern-character as 
\be 
\text{Ch}(a)=f\wedge f=da\wedge da=d(a\wedge da).
\ee Since the Chern-character is an exact form locally, we can use Stokes' theorem 
\be 
\int_{\mathcal M} \text{Ch}(a)=\int_{t=t_f} a\wedge da-\int_{t=t_i}a\wedge da+\int_{r=R}a\wedge da,\label{stokes}
\ee 
where the four-dimensional manifold $\mathcal M$ is bounded by two spacelike hypersurfaces $t=t_{i/f}$ and a timelike hypersurface $r=R$. This is shown in Figure \ref{fig:rectangle_region}. We can take the limit  $R\to \infty$ and  $t_i\to-\infty, t_f\to\infty$ such that $\mathcal{M}$ approaches the whole spacetime $\mathbb R^{1,3}$. 
The first term corresponds to the magnetic helicity at future timelike infinity ($i^+$)
\be
\mathcal H_m(i^+)=\lim_{t_f\to\infty}\int_{t=t_f} a\wedge da
\ee 
while the second term corresponds to the magnetic helicity at past timelike infinity ($i^-$)
\be 
\mathcal H_m(i^-)=\lim_{t_i\to-\infty}\int_{t=t_i} a\wedge da.
\ee The third term should be evaluated at the spatial infinity ($i^0$)
\be 
H(i^0)=\int_{i^0}a\wedge da.
\ee 
Collecting all these results, we find the identity
\be 
\int_{\mathbb R^{1,3}}f\wedge f=\mathcal H_m(i^+)-\mathcal H_m(i^-)+H(i^0).\label{fwedgef}
\ee 
One can also consider a causal diamond with radius $R$ and then take the limit $R\to \infty$ to cover the whole spacetime. This  is shown in Figure \ref{fig:diamond1_coords}. Using Stokes' theorem, we should find 
\bea 
\int_{\mathbb R^{1,3}} f\wedge f=H(\mathcal I^+)-H(\mathcal I^-)
\eea  where $H(\mathcal I^+)$ is exactly the radiative helicity that crosses $\mathcal I^+$. Similar definition applies to $H(\mathcal I^-)$. 

In Appendix \ref{magsupp}, we will discuss the differences and connections between magnetic helicity and radiative helicity via more examples. 

Besides magnetic helicity, one can use the duality invariance of vacuum Maxwell equation to define the electric helicity \cite{Trueba1996TheEH} 
\be 
\mathcal{H}_{e}=\int_V d^3\bm x \tilde{\bm a}\cdot\bm e
\ee where $\tilde{\bm a}$ is the spatial component of the dual vector potential $\tilde{a}$ associated with the Hodge dual of $f_{\mu\nu}$
\be 
d\tilde{a}=* f.
\ee Here $\bm e$ is the electric field. Similar to the magnetic helicity, the electric helicity is obviously not the radiative helicity studied in this work.
\subsubsection{Optical helicity}
\label{optical}
In literature, there is also a quantity which is the combination of the magnetic and electric helicity 
\be 
\mathcal{H}_{\text{op}}=\frac{1}{2}\int d^3\bm x(\bm a\cdot\bm b-\tilde{\bm a}\cdot\bm e).
\ee  We will follow \cite{optical} to call it optical helicity,   although it was discovered  sixty years ago \cite{newconservation}. The optical helicity is the difference of photon numbers of opposite helicity as a consequence of the duality invariance of the vacuum Maxwell theory. Both of these properties match the radiative helicity $H$ in this work. However, we will focus on the helicity flux density $O(u,\Omega)$ which is defined at $\mathcal I^+$ while the optical helicity density \cite{optical}
\be 
O_{\text{op}}(\bm x)=\frac{1}{2}(\bm a\cdot\bm b-\tilde{\bm a}\cdot\bm e)
\ee is defined on the spacelike hypersurface $V$.
Obviously, these two densities are not  equivalent \be 
O(u,\Omega)\not=O_{\text{op}}(\bm x).
\ee The former is only related to the radiative field near $\mathcal I^+$, whereas the latter is determined by the field configuration deep in the bulk. Readers can also convince themselves that the latter is gauge dependent in general. However, the helicity flux density is  gauge invariant, up to possible large gauge transformations. 

Recall that after fixing the gauge, there are still residual gauge transformations  among which the large gauge transformation preserves the fall-off condition and leads to non-trivial charges. It can be shown that the large gauge transformation is generated by a time-independent function $\eta(\Omega)$
\be 
\delta_\eta A_A(u,\Omega)=\partial_A \eta(\Omega)\quad\Rightarrow\quad \delta_\eta\dot A_A(u,\Omega)=0.
\ee Therefore, the energy flux density is invariant under the large gauge transformation
\be 
\delta_\eta T(u,\Omega)=0
\ee while the angular momentum/helicity flux density is invariant up to a full derivative term 
\bea 
\delta_\eta M_{A}(u,\Omega)&=& \frac{d}{du}[\frac{1}{2}(A^B\nabla^C\nabla^D\eta(\Omega)-\nabla^B\eta(\Omega)\nabla^CA^D)P_{ABCD}],\\
\delta_\eta O(u,\Omega)&=&\frac{d}{du}[\epsilon_{AB}A^A(u,\Omega)\partial^B\eta(\Omega)].
\eea It makes sense that the angular momentum/helicity flux changes under large gauge transformation since this transformation is physical. As an example, the angular momentum in an asymptotically flat spacetime suffers supertranslation ambiguity \cite{Penrose+1982+631+668}. This ambiguity has been a persistent theoretical challenge for decades \cite{Ashtekar:1981bq,PhysRevLett.81.1150,Hawking_2017}, and recent studies have introduced correction terms aimed at resolving it in gravitational theory \cite{chen2021supertranslationinvarianceangularmomentum,Mao_2023,javadinezhad2024puzzlescovariancesupertranslationinvariance}.
Ambiguity of the angular momentum caused by supertranslation at higher-order perturbative levels and the global consistency of gauge fixing remain a subject of ongoing investigation \cite{Veneziano_2025}. Similarly, the helicity density suffers an ambiguity under the large gauge transformation
\be 
\Delta_\eta\left(\frac{dH}{d\Omega}\right)=\int_{u_i}^{u_f} du \delta_\eta O(u,\Omega)=\epsilon_{AB}[A^A(u_f,\Omega)-A^A(u_i,\Omega)]\partial^B \eta(\Omega).
\ee 
Further conclusions can be drawn as follows:
\begin{itemize}
    \item Assume $u_i\to-\infty$ and $u_f\to \infty$, the variation of the large gauge transformation is 
    \be 
    \Delta_\eta \left(\frac{dH}{d\Omega}\right)=\epsilon_{AB}[A^A(\infty ,\Omega)-A^A(-\infty,\Omega)]\partial^B \eta(\Omega)
    \ee where $A^A(\infty ,\Omega)-A^A(-\infty,\Omega)$ is exactly the soft mode in \cite{Strominger:2017zoo}.
    \item For a system with period $T$, then the  helicity density is invariant  under the large gauge transformation during a period ($u_f=u_i+T$)
    \be 
    \Delta_\eta\left(\frac{dH}{d\Omega}\right)=\epsilon_{AB}[A^A(u_i+T ,\Omega)-A^A(u_i,\Omega)]\partial^B \eta(\Omega)=0.
    \ee 
\end{itemize}

\subsubsection{Chern number}
In literature, an approach from topology and anomaly is closely related to our work. In this part, we will briefly discuss the connections and differences between the topological approach and our paper. Usually,  
an anomaly is the breakdown  of classical symmetry after 
quantization. In history, the ABJ anomaly \cite{Adler:1969gk,Bell:1969ts} promoted the understanding of the relationship between anomaly and topology. In \cite{Liu:2024rvz, Guo:2024qzv}, the connection between the radiative helicity and topological Chern number has been discussed in detail. In other words, one  evaluates the  Chern-Simons term at $\mathcal I^+$ and finds  the radiative helicity
\be 
H=\int_{\mathcal I^+}a\wedge da=\int du d\Omega \epsilon_{AB}A^A\dot A^B.
\ee  However, in our work, the helicity flux density $O(u,\Omega)$ itself is not a topological invariant. One should integrate it over $\mathcal I^+$ to obtain the radiative helicity (topological term). More precisely, by introducing a test function $g(\Omega)$, one can define the so-called helicity flux operator \cite{Liu:2023qtr}
\be 
\mathcal O_g=\int du d\Omega g(\Omega)O(u,\Omega)
\ee that represents the angle distribution of the helicity flux density. When $g=1$, the operator $\mathcal O_{g=1}$ is identified as the topological  quantity. However, one can extract more information by choosing $g(\Omega)$ freely. In general, $g(\Omega)$ is decomposed into linear superposition of spherical harmonics 
\be 
g(\Omega)=\sum_{\ell=0}^\infty\sum_{m=-\ell}^\ell g_{\ell,m} Y_{\ell,m}(\Omega)
\ee and then the operator $\mathcal O_g$ is decomposed into an infinite set of observables 
\be 
O_{\ell,m}=\int du d\Omega Y_{\ell,m}(\Omega)O(u,\Omega).
\ee The main point in our work is that the helicity flux density itself is an observable, besides the topological quantity.

\section{EM helicity flux density}\label{ems}
In this section, we will derive the EM helicity flux density caused by the sources. 
\subsection{General current}
The vector potential generated by a general charge current is  
\be 
a_\mu(t,\bm x)=\frac{1}{4\pi}\int_V d^3\bm x' \frac{j_\mu(t-|\bm x-\bm x'|,\bm x')}{|\bm x-\bm x'|}.
\ee 
Near $\mathcal I^+$, we use the following expansion 
\be 
\frac{1}{|\bm x-\bm x'|}=\frac{1}{r}+\mathcal{O}(r^{-2}),\quad t-|\bm x-\bm x'|=u+\bm n\cdot\bm x'+\mathcal{O}(r^{-1})
\ee and then 
\be 
A_\mu(u,\Omega)=\lim_{r\to\infty,\ u\ \text{finite}}\ r\times  a_\mu(t,\bm x)=\frac{1}{4\pi}\int_V d^3\bm x' j_\mu(u+\bm n\cdot\bm x',\bm x').
\ee As a consequence, the transverse modes at $\mathcal I^+$ are
\be 
A_A(u,\Omega)=-Y_A^i A_i(u,\Omega)=-\frac{Y^i_A}{4\pi}\int_V d^3\bm x' j_i(u+\bm n\cdot\bm x',\bm x')
\label{aA}
\ee
where the vector field in the spatial direction is 
\be 
A_i(u,\Omega)=\frac{1}{4\pi}\int_V d^3\bm x' j_i(u+\bm n\cdot\bm x',\bm x').\label{Aij}
\ee The energy and helicity flux densities are 
\begin{align}
    \frac{dE}{du d\Omega}=T(u,\Omega)&=P_{ij}\dot{A}_i\dot A_j=\dot{\bm A}\cdot\dot{\bm A}-(\bm n\cdot\dot{\bm A})^2,\label{energyflux1}\\
   \frac{dH}{du d\Omega}= O(u,\Omega)&=\epsilon_{ijk}n_i A_j\dot A_k={\bm n\cdot(\bm A\times\dot{\bm A})}\label{helicityflux1}
\end{align} where we have used the following identities 
\be 
Y_i^A Y_{jA}=P_{ij}=\delta_{ij}-n_i n_j,\quad \epsilon_{AB}Y^A_i Y^B_j=\epsilon_{ijk}n_k.
\ee 
Switching to the frequency space, 
\be 
A_i(\omega,\Omega)=\int du A_i(u,\Omega)e^{-i\omega u},\quad A_i(u,\Omega)=\frac{1}{2\pi}\int d\omega A_i(\omega,\Omega)e^{i\omega u},
\ee we find 
\bea 
\frac{dE}{d\omega d\Omega}&=&\frac{1}{\pi}\omega^2 P_{ij}A_i(\omega,\Omega)A_j^*(\omega,\Omega),\label{domegaE}\\
\frac{dH}{d\omega d\Omega}&=&-\frac{i}{\pi}\omega n_i\epsilon_{ijk}A_j(\omega,\Omega)A_k^*(\omega,\Omega).\label{domegaH}
\eea 
Note that the field $A_i(u,\Omega)$ is real and thus \be 
A_i^*(\omega,\Omega)=A_i(-\omega,\Omega).
\ee Therefore, we have multiplied a factor 2 on the right hand side of \eqref{domegaE} and \eqref{domegaH} since only the positive modes are independent.
\subsection{Point particle}
For a point particle with charge $q$ whose trajectory is parameterized by $ 
\bm x_s(t)
$, the four-current is 
\be 
j^\mu(t,\bm x)=q (1,\bm v(t))\delta^{(3)}(\bm x-\bm x_s(t))
\ee where $\bm v(t)=\frac{d}{dt}\bm x_s(t)$ is the three-velocity of the particle. Substituting it into \eqref{Aij}, we obtain 
\be 
A_i(u,\Omega)=\frac{1}{4\pi}\int_V d^3\bm x' q\bm v(u+\bm n\cdot\bm x')\delta^{(3)}(\bm x'-\bm x_s(u+\bm n\cdot\bm x'))=\frac{1}{4\pi}\frac{qv_i}{1-\bm n\cdot\bm v}\Big|_{t_r}
\ee where $t_r$  in the direction of $\bm n$ is 
\be 
t_r=u+\bm n\cdot\bm x_s(t_r).\label{tr}
\ee Though $t_r$ is also called retarded time in \cite{jackson1999classical}, it is not equal to the retarded time $u=t-r$ in our paper. When the size of the source is small, they are  equivalent to each other approximately through equation \eqref{tr}. Taking the derivative  with respect to $u$, we find 
\be 
\frac{dt_r}{du}=\frac{1}{1-\bm n\cdot\bm v}\Big|_{t_r}.\label{dtr}
\ee Therefore, 
\be 
\dot A_i(u,\Omega)=\frac{d}{du}A_i(u,\Omega)=\frac{1}{4\pi}\frac{q}{(1-\bm n\cdot\bm v)^2}[a_i+\frac{v_i(\bm n\cdot \bm a)}{1-\bm n\cdot\bm v} ]\Big|_{t_r}
\ee where the acceleration $\bm a$ is defined as \footnote{Unfortunately, the same symbol $\bm a$ also refers to the vector potential in previous sections. We apologize for such abuse of symbol. } 
\be 
\bm a(t)=\frac{d}{dt}\bm v(t).
\ee Then the energy and helicity flux densities are 
\bea 
\frac{dE}{du d\Omega}&=&\frac{q^2}{16\pi^2}\frac{|\bm n\times\left((\bm n-\bm v)\times\bm a\right)|^2}{(1-\bm n\cdot\bm v)^6},\\
\frac{dH}{du d\Omega}&=&\frac{q^2}{16\pi^2}\frac{\bm n\cdot(\bm v\times\bm a)}{(1-\bm n\cdot\bm v)^3}.
\label{monop}
\eea Note that on the right hand side, the velocity and the acceleration are evaluated at the time $t_r$. One may use the identity \eqref{dtr} to rewrite them as 
\bea 
\frac{dE}{dt_r d\Omega}&=&\frac{q^2}{16\pi^2}\frac{|\bm n\times\left((\bm n-\bm v)\times\bm a\right)|^2}{(1-\bm n\cdot\bm v)^5},\label{dE}\\
\frac{dH}{dt_rd\Omega}&=&\frac{q^2}{16\pi^2}\frac{\bm n\cdot(\bm v\times\bm a)}{(1-\bm n\cdot\bm v)^2}.\label{dH}
\eea  The modified energy flux density \eqref{dE} matches  the one in \cite{jackson1999classical}. The equation \eqref{dH} has not appeared in the literature as far as we know. Integrating over the sphere, we find the energy/helicity flux 
\bea 
\frac{dE}{du}&=&\frac{q^2}{16\pi^2}\int d\Omega \frac{|\bm n\times\left((\bm n-\bm v)\times\bm a\right)|^2}{(1-\bm n\cdot\bm v)^6},\label{intde}\\
\frac{dH}{du}&=&\frac{q^2}{16\pi^2}\int d\Omega \frac{\bm n\cdot(\bm v\times\bm a)}{(1-\bm n\cdot\bm v)^3}.\label{intdh}
\eea The argument in the velocity and the acceleration is $t_r=u+\bm n\cdot\bm x_s(t_r)$, we can solve it implicitly as
\be 
t_r=t_r(u,\bm n),
\ee the integrals  \eqref{intde} and \eqref{intdh} can not be worked out explicitly. However, one can define the modified energy/helicity flux 
\bea 
\frac{dE}{dt_r}&=&\frac{q^2}{16\pi^2}\int d\Omega \frac{|\bm n\times\left((\bm n-\bm v)\times\bm a\right)|^2}{(1-\bm n\cdot\bm v)^5},\\
\frac{dH}{dt_r}&=&\frac{q^2}{16\pi^2}\int d\Omega \frac{\bm n\cdot(\bm v\times\bm a)}{(1-\bm n\cdot\bm v)^2}.
\eea After some efforts, the modified energy/helicity flux can be found explicitly 
\bea 
\frac{dE}{dt_r}&=&\frac{q^2}{6\pi}\gamma^6(\bm a_{\parallel}^2+\gamma^{-2}\bm a_{\perp}^2),\quad \gamma=(1-\bm v^2)^{-1/2},\label{detr}\\
\frac{dH}{dt_r}&=&0,\label{dhtr}
\eea  where the acceleration $\bm a$ is decomposed into the component $ \bm a_{\parallel}$ that is parallel to the 3-velocity $\bm v$, and the component $\bm a_{\perp}$ perpendicular to  $\bm v$. Again, the formula \eqref{detr} is exactly the total power radiated and reduces to the Larmor's formula in the non-relativistic limit  \cite{jackson1999classical}. The vanishing of the modified helicity flux follows from the fact that  the integral 
\be 
\int d\Omega \frac{n_i}{(1-\bm n\cdot\bm v)^2}
\ee is always proportional to the velocity and ${\bm v \cdot(\bm v\times\bm a)=0}$. Though it vanishes, the helicity flux density is still non-zero. Similar phenomenon has also been noticed for gravitational helicity flux density \cite{Long:2024yvj}. 

\paragraph{Remark.} The distinction between $t_r$ and $u$ is important since it leads to a factor discrepancy between the energy/helicity flux and the modified energy/helicity flux.
It is the modified energy/helicity flux density that matches  the definition in the textbook \cite{jackson1999classical}. However, there is no problem to consider the energy and helicity flux since they are the summation over the corresponding fluxes at an angle-independent time $u$. Moreover, the energy/helicity density can be obtained either from the energy/helicity flux density or from the modified one
\bea 
\frac{dE}{d\Omega}&=&\int du \frac{dE}{du d\Omega}=\int dt_r \frac{dE}{dt_r d\Omega},\\
\frac{dH}{d\Omega}&=&\int du \frac{dH}{du d\Omega}=\int dt_r \frac{dH}{dt_r d\Omega}.
\eea In the following, we will only consider the energy/helicity flux density.

\subsection{General properties}

Before delving into the toy models, we will  establish the general properties of the helicity flux density. 
\begin{enumerate}
\item For a point charge moving along a straight line, the direction of the velocity is always parallel to the acceleration and then the helicity flux density is zero 
\be 
\frac{dH}{du d\Omega}=0,\quad \bm v\parallel \bm a.
\ee {Recall that the helicity flux density measures the angular imbalance between left- and right-handed photons. Such an imbalance must originate from a source whose motion exhibits a handedness. For a point particle, this instantaneous handedness of its trajectory is  characterized by $\bm v\times\bm a$ term (see \eqref{monop}). In the case of rectilinear motion, $\bm v\parallel\bm a$ and the kinematic twist is zero. The system is axially symmetric and lacks any intrinsic handedness. Therefore, the source radiates an exactly equal number of left- and right-handed photons in all directions, resulting in a zero net helicity flux density everywhere.}

\item When the velocity and the acceleration are not collinear, they should generate a plane $\Pi(\bm v, \bm a)$.  Helicity flux density vanishes as observation directions $\bm n$ lying within this plane
\be
\frac{dH}{dud\Omega}=0,\quad \bm n\in\Pi(\bm v,\bm a).
\ee
{Unlike the case of rectilinear motion, the system here does possess handedness of trajectory, defined by a non-vanishing vector $\bm v\times \bm a$, which is orthogonal to the  plane $\Pi(\bm v,\bm a)$. The helicity flux density $O(u,\Omega)$ measures the projection of  this trajectory handedness onto the observer's line of sight $\bm n$. If the observer lies within the plane $\Pi(\bm v,\bm a)$,  then by definition the line of sight $\bm n$ is orthogonal to  $\bm v\times \bm a$. From this point of view, the source's handedness becomes invisible to the observer. The radiation appears chirally symmetric, with the left- and right-handed photon contributions exactly canceling, leading to zero net helicity flux.
}

\item In the non-relativistic limit, the helicity flux density is 
    \be 
    \frac{dH}{dud\Omega}=\frac{q^2}{16\pi^2}\bm n\cdot(\bm v\times\bm a).\label{NRhe}
    \ee {The formula, $\frac{dH}{dud\Omega}\propto \bm n\cdot(\bm v\times\bm a)$, states that the observed helicity flux density is proportional to the projection of the source's handedness $\bm v\times\bm a$ onto the observer's line of sight $\bm n$. The sign of this projection determines which helicity (left- or right-handed) is dominant in that direction, while its magnitude determines the degree of that dominance. This expression provides a unified explanation for the first two properties: the flux vanishes either when the source has no handedness ($\bm v\times\bm a=0$) or when the observer is in a geometric blind point ($\bm n \cdot(\bm v\times\bm a)=0$). Interestingly, the quantity $\bm v\times\bm a$ itself has, to our knowledge, rarely been emphasized in classical mechanics\footnote{ In special relativity, the same factor $\bm v\times\bm a$ appears in Thomas precession \cite{jackson1999classical}.}.  A general curvilinear motion in three dimensions can be described by the Frenet-Serret formulae
\begin{align}
\frac{d\boldsymbol{t}}{ds}&=\kappa \boldsymbol{n},\nn
\\
\frac{d\boldsymbol{n}}{ds}&=\tau \boldsymbol{b}-\kappa \boldsymbol{t},\label{FSform}
\\
\frac{d\boldsymbol{b}}{ds}&=-\tau \boldsymbol{n}.\nn
\end{align} Here we use the traditional conventions consistent with the usual differential geometry text, where the parameter $s$ is the arc length of the  curve, the right-handed frame bases $\bm t,\bm n,\bm b$ are tangent vector, principal normal vector and binormal vector of the curve respectively, and $\kappa,\tau$ are curvature and torsion of the curve respectively. It can be shown that the magnitude $|\bm v\times\bm a|$ is proportional to the curvature $\kappa$ \cite{banchoff2016differential}
\be 
|\bm v\times\bm a|=|\bm v|^3 \kappa
\ee and its direction coincides with  the binormal vector $\bm b$
\be \bm b=\frac{\bm v\times\bm a}{|\bm v\times\bm a|}.
\ee Consequently, the handedness of the trajectory admits a clear geometric interpretation
\be 
\bm v\times\bm a=|\bm v|^3\kappa\bm b.
\ee A further exploration of this quantity is left for future work.

}

    \item For $N$ moving charged particles, the helicity flux density is 
    \be
    \frac{dH_{1,2,\cdots,N}}{du d\Omega}=\frac{1}{16\pi^2}\sum_{I,J=1}^Nq_Iq_J[\frac{\bm n\cdot(\bm v^{(I)}\times\bm a^{(J)})}{(1-\bm n\cdot \bm v^{(I)})(1-\bm n\cdot \bm v^{(J)})^2}+\frac{(\bm n\cdot(\bm v^{(I)}\times\bm v^{(J)}))( \bm n\cdot \bm a^{(J)})}{(1-\bm n\cdot \bm v^{(I)})(1-\bm n\cdot \bm v^{(J)})^3}]
    \ee where $q_I$ is the charge of the $I$-th particle. The velocity and the acceleration of $I$-th particle are labeled as $\bm v^{(I)}$ and $\bm a^{(I)}$ respectively. There is an interference effect due to the mixing terms
    \bea 
   && \frac{dH_{1,2,\cdots,N}}{dud\Omega}-\sum_{I=1}^N\frac{dH_I}{du d\Omega}\nn\\&=&\frac{1}{16\pi^2}\sum_{I\not=J}q_Iq_J[\frac{\bm n\cdot(\bm v^{(I)}\times\bm a^{(J)})}{(1-\bm n\cdot \bm v^{(I)})(1-\bm n\cdot \bm v^{(J)})^2}+\frac{(\bm n\cdot(\bm v^{(I)}\times\bm v^{(J)}))( \bm n\cdot \bm a^{(J)})}{(1-\bm n\cdot \bm v^{(I)})(1-\bm n\cdot \bm v^{(J)})^3}].\nn\\
    \eea This becomes rather clear for two particles. The interference effect for the helicity flux density is 
    \bea 
    &&\frac{dH_{1,2}}{du d\Omega}-\frac{dH_1}{du d\Omega}-\frac{dH_2}{du d\Omega}\nn\\&=&\frac{q_1q_2}{16\pi^2}[\frac{\bm n\cdot(\bm v^{(1)}\times\bm a^{(2)})}{(1-\bm n\cdot \bm v^{(1)})(1-\bm n\cdot \bm v^{(2)})^2}+\frac{\bm n\cdot(\bm v^{(2)}\times\bm a^{(1)})}{(1-\bm n\cdot \bm v^{(2)})(1-\bm n\cdot \bm v^{(1)})^2}\nn\\&&+\frac{(\bm n\cdot(\bm v^{(1)}\times\bm v^{(2)}))( \bm n\cdot \bm a^{(2)})}{(1-\bm n\cdot \bm v^{(1)})(1-\bm n\cdot \bm v^{(2)})^3}+\frac{(\bm n\cdot(\bm v^{(2)}\times\bm v^{(1)}))( \bm n\cdot \bm a^{(1)})}{(1-\bm n\cdot \bm v^{(2)})(1-\bm n\cdot \bm v^{(1)})^3}].
    \label{h12}
    \eea In magnetic helicity, similar terms are called mutual magnetic helicity. Several properties of the mutual magnetic helicity can be found in \cite{2006SoPh..233....3D}. We therefore designate these interference terms as mutual helicity flux density.  {Physically, these interference terms arise as cross terms from the superposition of the individual radiation fields in the quadratic expression for the helicity flux. Mutual magnetic helicity is a pure geometric quantity that qualifies the relative winding and linkage of distinct magnetic flux tubes\cite{2006SoPh..233....3D}. Mutual helicity flux density can therefore be physically interpreted as the radiative consequence of a kinematic winding between the particles' trajectories. The term in \eqref{h12}, such as $\bm n\cdot(\bm{v}^{(1)}\times\bm{a}^{(2)})$, describes how the velocity of one particle and the acceleration of another jointly create a ``mutual handedness''. This kinematic winding acts as a combined, coherent source that radiates a net imbalance of left- and right-handed photons along the direction $\bm{n}$.}

\end{enumerate}
\subsection{Toy models}
In this subsection, we will consider four toy models to study the angular distribution of the EM helicity flux density. 
\subsubsection{Circular motion}
\begin{figure}[h!]
    \centering
    \includegraphics[width=0.9\textwidth]{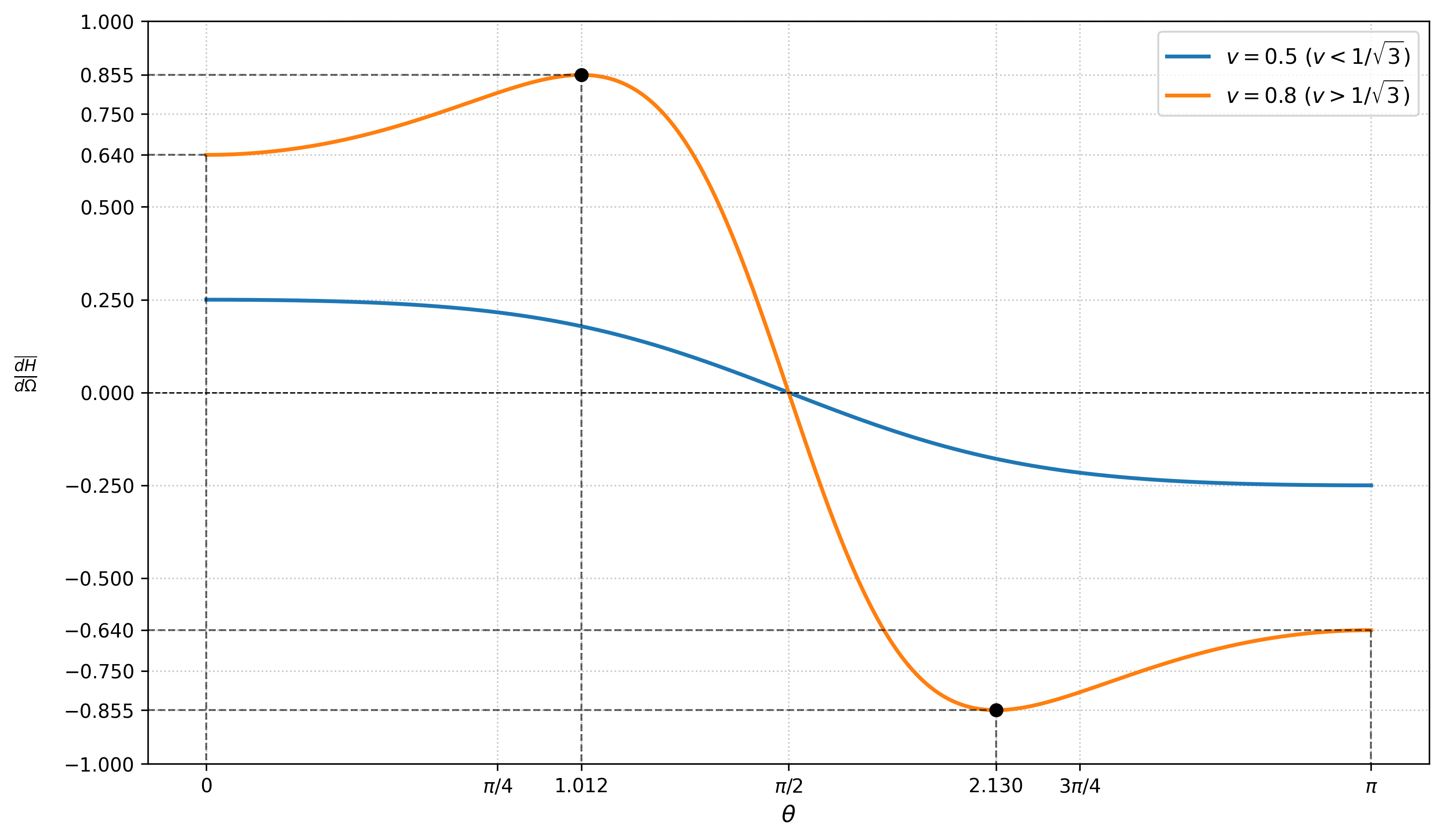}
    \caption{
        Angular distribution of the rescaled helicity density $\frac{\overline{dH}}{d\Omega}$ for a circular motion. 
    }
    \label{fig:circular_motion_helicity}
\end{figure}

For a particle with charge $q$ that is moving in a circle with radius $R$ in $x-y$ plane with frequency $\omega$ \footnote{One may assume the particle moves in the plane perpendicular to a uniform magnetic field.}, the trajectory of the particle is 
\be 
\bm x_s(t)=(R\cos\omega t, R\sin\omega t,0).
\ee The velocity and the acceleration are 
\bea 
\bm v(t)&=&R\omega(-\sin\omega t,\cos\omega t,0),\\
\bm a(t)&=&-R\omega^2(\cos\omega t,\sin\omega t,0).
\eea 
The velocity of the particle should be less than the velocity of light due to the constraint from relativity, it follows that 
\be 
0< v=R\omega<1.
\ee 
The energy and helicity flux density are 
\bea 
\frac{dE}{du d\Omega}&=&\frac{q^2v^2\omega^2}{128 \pi ^2}\frac{-4 \sin \theta  (\sin \theta  \cos (2 (\omega t_r-\phi ))-4 v \sin (\omega t_r-\phi ))+\left(2-4 v^2\right) \cos (2 \theta )+4 v^2+6}{\left(1+v \sin\theta \sin(\omega t_r-\phi)\right)^6},\\
\frac{dH}{du d\Omega}&=&\frac{q^2}{16\pi^2}\frac{v^2\omega\cos\theta}{\left(1+v \sin\theta \sin(\omega t_r-\phi)\right)^3},
\eea where $t_r$ is related to $u$ by the equation 
\be 
t_r=u+R\sin\theta \cos(\omega t_r-\phi).
\ee The energy and helicity density during a period are
\bea 
\frac{dE}{d\Omega}&=&\int_{u_i}^{u_f} \frac{dE}{du d\Omega}du=\int_{-\pi/\omega}^{\pi/\omega}\frac{dE}{du d\Omega}\frac{du}{dt_r} dt_r=\frac{q^2 v^2 \omega  \left(8- \left(3 v^4+v^2\right)\sin^4\theta-4\sin^2\theta\right)}{64 \pi  \left(1- v^2\sin^2\theta\right)^{7/2}},\\
\frac{dH}{d\Omega}&=&\int_{u_i}^{u_f} \frac{dH}{du d\Omega}du=\int_{-\pi/\omega}^{\pi/\omega}\frac{dH}{du d\Omega}\frac{du}{dt_r} dt_r=\frac{q^2v^2\cos\theta}{8\pi(1-v^2\sin^2\theta)^{3/2}}\label{circularhelicity}
\eea 
where we have used the integral identity 
\be 
\int_{-\pi}^\pi \frac{dt}{(1+v\sin t)^\alpha}=2 \pi  \, _2F_1\left(\frac{\alpha}{2},\frac{\alpha+1}{2};1;v^2\right).
\ee For integer $\alpha$, the above hypergeometric function becomes elementary function. By integrating over the solid angle, we find the radiative energy and helicity during a period
\bea 
E&=&\frac{q^2 v^2 \omega}{3(1-v^2)^2},\label{radiativeenergy}\\
H&=&0.
\eea The average radiant power is the radiative energy divided by the period $T=\frac{2\pi}{\omega}$
\be 
\frac{E}{T}=\frac{q^2a^2}{6\pi(1-v^2)^2},\label{radiativepower}
\ee where $a=v\omega$ is the magnitude of the acceleration. The radiant power \eqref{radiativepower} matches  the general result \eqref{detr}. The radiative helicity is zero, consistent with the general argument \eqref{dhtr}. Interestingly, the helicity density is still non-vanishing. Several properties are discussed in the following: 
\begin{enumerate}
    \item The helicity density  is parity odd. By changing $\theta\to\pi-\theta$, the sign of the helicity density is flipped.
    \item The extreme value of the helicity density depends on the magnitude of the velocity, it is located at the angle 
    \be 
    \cos\theta_*=\pm\sqrt{\frac{1-v^2}{2v^2}}\label{thetastar}
    \ee 
    when $\frac{1}{\sqrt{3}}<v<1$. There are real solutions for \eqref{thetastar}. In this case, the maximal value of the helicity density is located at 
    \be 
    \theta_*=\text{arccos}\sqrt{\frac{1-v^2}{2v^2}}\label{cos}
    \ee with 
    \bea 
    \frac{dH}{d\Omega}\Big|_{\text{max}}=\frac{q^2 v}{12\sqrt{3}\pi(1-v^2)}.\label{maxvalue}
    \eea Similarly, the minimal value of the helicity density is located at 
$\pi-\theta_*$ with 
\bea 
\frac{dH}{d\Omega}\Big|_{\text{min}}=-\frac{dH}{d\Omega}\Big|_{\text{max}}.
\eea
    When $0<v<\frac{1}{\sqrt{3}}$, there is no real solution for equation \eqref{thetastar}. As a consequence, the maximal/minimal value of the helicity density is located at the north/south pole $\theta=0\ \text{or}\ \pi$
    \be 
    \frac{dH}{d\Omega}\Big|_{\text{max}}=\frac{q^2v^2}{8\pi},\quad  \frac{dH}{d\Omega}\Big|_{\text{min}}=-\frac{q^2v^2}{8\pi}.
    \ee  
\end{enumerate}
{In Figure \ref{fig:circular_motion_helicity}, we plot the angular distribution of the rescaled helicity density 
\be 
\frac{\overline {dH}}{d\Omega}=\frac{v^2\cos\theta}{(1-v^2\sin^2\theta)^{3/2}}
\ee for a charged particle in circular motion. The helicity density is a function of the polar angle $\theta$. We have chosen two different velocities, the blue curve corresponds to  $v=0.5<1/\sqrt{3}$ and the red one corresponds to $v=0.8> 1/\sqrt{3}$. The maximum/minimum value of the helicity density is indeed located the north/south pole for $v=0.5$. On the other hand, when $v=0.8$, the maximum value is located at $\theta_*\approx 1.012$ and the minimum value is located at $\pi-\theta_*\approx2.130$. The location is consistent with the formula \eqref{cos}. We have also checked the formula \eqref{maxvalue} for the maximum value.
}

\subsubsection{Helical motion}
In this case, the charged particle moves in a uniform magnetic field, but its initial velocity is not perpendicular to the magnetic field. The trajectory of the particle is 
\bea 
\bm x_s(t)=R \cos\omega t\bm e_x+R \sin\omega t \bm e_y+v_{\parallel} t \bm e_z, 
\eea where $v_{\parallel}$ is the component of the velocity that is parallel to the magnetic field lines. Thus the velocity and the acceleration of the particle are 
\begin{align}
    \bm v(t)&=-R\omega \sin\omega t \bm e_x+R\omega \cos\omega t\bm e_y+ v_{\parallel}\bm e_z,\\
    \bm a(t)&=-R\omega^2 (\cos\omega t\bm e_x+\sin\omega t \bm e_y).
\end{align} We set $v_{\perp}=R\omega$ to be the magnitude of the velocity perpendicular to the $x-y$ plane and then the helicity flux density is 
\be 
\frac{dH}{du d\Omega}=\frac{q^2v_{\perp}\omega }{16\pi^2}\frac{v_{\perp}\cos\theta+v_{\parallel}\sin\theta \sin(\omega t_r-\phi)}{(1+v_{\perp}\sin\theta \sin(\omega t_r-\phi)-v_{\parallel}\cos\theta)^3}.
\ee Therefore, the helicity density in a period is 
\bea 
\frac{dH}{d\Omega}=\frac{q^2v_{\perp}^2}{8\pi}\frac{\cos\theta-v_{\parallel}}{\left((1-v_{\parallel}\cos\theta)^2-v_{\perp}^2\sin^2\theta\right)^{3/2}}.
\eea In the limit $v_{\parallel}\to 0$, the particle moves in circular orbit and the helicity density reduces to \eqref{circularhelicity}. We can also find the following parity symmetry of the helicity density 
\be 
\frac{dH}{d\Omega}(v_{\perp},v_{\parallel},\theta)=-\frac{dH}{d\Omega}(v_{\perp},-v_{\parallel},\pi-\theta).
\ee We will assume $v_{\parallel}>0$ and denote the magnitude of the velocity by $v=\sqrt{v_{\perp}^2+v_{\parallel}^2}$. Then the extreme value of the helicity density is determined by the quadratic equation 
\be 
P(v,v_{\parallel},\lambda)\equiv 2v^2 \lambda^2-v_{\parallel}(1+3v^2)\lambda-(1-v^2-2v_{\parallel}^2)=0\label{lambdaeq}
\ee where $-1\le \lambda=\cos\theta\le 1$ and the discriminant is 
\be 
\text{dis}(v,v_{\parallel})=-(1-v^2)\left((9v^2-1)v_{\parallel}^2-8v^2\right).\label{ineqv}
\ee 
The minimal value of the quadratic equation \eqref{lambdaeq} is located at 
\be 
\lambda_{\text{min}}=\frac{v_{\parallel}(1+3v^2)}{4v^2}
\ee with 
\be 
P(v,v_{\parallel},\lambda_{\text{min}})=-\frac{\text{dis}(v,v_{\parallel})}{8v^2}.
\ee 
In Appendix \ref{min}, we prove that the discriminant is always positive for any $0<v<1$. Therefore, the two solutions of \eqref{lambdaeq} are real
\be
\lambda_{\pm}=\frac{1}{4v^2}\left[v_{\parallel}(1+3v^2)\pm\sqrt{\text{dis}(v,v_{\parallel})}\right].
\ee 
We can also find 
\begin{align} 
P(v,v_{\parallel},-1)&=(1+v_{\parallel})(2v_{\parallel}+3v^2-1)=(1+v_{\parallel})f_-(v,v_{\parallel}),\\                                                                                                                                                                                                                                                                                                                                                                                                                                                                                                                                                                                                                                                                                                                                                                                                                                                                                                                                                                                                                                                                                                                                                                                                                                                                                                                                                                                                                                                                                                                                                                                                                                                                                                                                                                                                                                                                                                                                                                                                                                                                                                                                                                                                                                                                                                                                                                                                                                                                                                                                                                                                                                                                                                                                                                                                                                                                                                                                                                                                                                                                                                                                                                                                                                                                                                                                                                                                                                                                                                                                                                                                                                                                                                                                                                                                                                                                                                                                                                                                                                                                                                                                                                                                                                                                                                                                                                                                                                                                                                                                                                                                                                                                                                                                                                                                                                                                                                                                                                                                                                                                                                                                                                                                                                                                                                                                                                                                                                                                                                                                                                                                                                                                                                                                                                                                                                                                                                                                                                                                                                                                                                                                                                                                                                                                                                                                                                                                                                                                                                                                                                                                                                                                                                                                                                                                                                                                                                                                                                                                                                                                                                                                                                                                                                                                                                                                                                                                                                                                                                                                                                                                                                                                                                                                                                                                                                                                                                                                                                                                                                                                                                                                                                                                                                                                                                                                                                                                                                                                                                                                                                                                                                                                                                                                                                                                                                                                                                                                                                                                                                                                                                                                                                                                                                                                                                                                                                                                                                                                                                                                                                                                                                                                                                                                                                                                                                                                                                                                                                                                                                                                                                                                                                                                                                                                                                                                                                                                                                                                                                                                                                                                                                                                                                                                                                                                                                                                                                                                                                                                                                                                                                                                                                                                                                                                                                                                                                                                                                                                                                                                                                                                                                                                                                                                                                                                                                                                                                                                                                                                                                                                                                                                                                                                                                                                                                                                                                                                                                                                                                                                                                                                                                                                                                                                                                                                                                                                                                                                                                                                                                                                                                                                                                                                                                                                                                                                                                                                                                                                                                                                                                                                                                                                                                                                                                                                                                                                                                                                                                                                                                                                                                                                                                                                                                                                                                                                                                                                                                                                                                                                                                                                                                                                                                                                                                                                                                                                                                                                                                                                                                                                                                                                                                                                                                                                                                                                                                                                                                                                                                                                                                                                                                                                                                                                                                                                                                                                                                                                                                                                                                                                                                                                                                                                                                                                                                                                                                                                                                                                                                                                                                                                                                                                                                                                                                                                                                                                                                                                                                                                                                                                                                                                                                                                                                                                                                                                                                                                                                                                                                                                                                                                                                                                                                                                                                                                                                                                                                                                                                                                                                                                                                                                                                                                                                                                                                                                                                                                                                                                                                                                                                                                                                                                                                                                                                                                                                                                                                                                                                                                                                                                                                                                                                                                                                                                                                                                                                                                                                                                                                                                                                                                                                                                                                                                                                                                                                                                                                                                                                                                                                                                                                                                                                                                                                                                                                                                                                                                                                                                                                                                                                                                                                                                                                                                                                                                                                                                                
P(v,v_{\parallel},1)&=(1-v_{\parallel})(3v^2-1-2v_{\parallel})=(1-v_{\parallel})f_+(v,v_{\parallel})
\end{align} where we have defined two functions $f_{\pm}(v,v_{\parallel})$ that determine the sign of $P(v,v_{\parallel},\pm)$.
In Figure \ref{phase}, we plot three curves in the $v_{\parallel}$-$v$ diagram,  the curve $v_{\parallel}=v$ and $P(v,v_{\parallel},-1)=0$ as well as $P(v,v_{\parallel},1)=0$. There are three regions that are relevant to our discussion.  We have labeled the intersection points as $o,\alpha,\beta,\gamma$ with the coordinates $(v,v_{\parallel})$
\begin{figure}[ht]
    \centering
\begin{tikzpicture}
    \begin{axis}[
        title={}, 
        title style={yshift=-5ex},
        xlabel={$v$}, 
        ylabel={$v_{\parallel}$}, 
        every axis x label/.style={at={(ticklabel* cs:1.00)},anchor=west, font=\small},
        every axis y label/.style={at={(ticklabel* cs:1.00)},anchor=south, font=\small},
        xmin=0, xmax=1.1,          
        ymin=0, ymax=1.2,      
        axis lines=left,         
        legend pos=outer north east, 
        domain=0:1,              
        samples=201,             
        smooth,                  
        legend cell align={left},
        xtick=\empty,
        ytick=\empty,
        clip mode=individual,
    ]

    \addplot[
        color=green,
        thick, 
        dashed,
    ] {x}; 
    \addlegendentry{$v_{\parallel} = v$};

    \addplot[
        name path=curveBlue,
        color=blue,
        thick,
        domain={1/3}:{1/sqrt(3)}, 
    ] {(1 - 3*x^2)/2};
    \addlegendentry{$v_{\parallel} = \frac{1-3v^2}{2}$}; 

    \addplot[
        color=red,
        thick,
    ] {(3*x^2 - 1)/2};
    \addlegendentry{$v_{\parallel} = \frac{3v^2-1}{2}$}; 

\draw[dashed, gray] (axis cs:1,0) -- (axis cs:1,1);

\path[name path=xaxis] (axis cs:0,0) -- (axis cs:1,0);
\node[text width=2cm, align=center] at (axis cs:0.3, 0.08) {{\Large I} \small $(-,-)$};
\node[text width=2cm, align=center] at (axis cs:0.6, 0.35) {{\Large II} \small $(-,+)$};
\node[text width=2cm, align=center] at (axis cs:0.88, 0.15) {{\Large III} \small $(+,+)$};

\node[below left, font=\small] at (axis cs:0,0) {o};
    \filldraw[black] (axis cs:0,0) circle (1.5pt);
\coordinate (alpha) at (axis cs:{1/3}, {1/3});
    \node[above right, font=\small] at (alpha) {$\alpha$};
    \filldraw[black] (alpha) circle (1.5pt);
 \coordinate (beta) at (axis cs:{1/sqrt(3)}, 0);
    \node[below right, font=\small] at (beta) {$\beta$};
    \filldraw[black] (beta) circle (1.5pt);
\coordinate (gamma) at (axis cs:1,1);
    \node[above left, font=\small] at (gamma) {$\gamma$};
    \filldraw[black] (gamma) circle (1.5pt);
\coordinate (delta) at (axis cs:1,0);
    \node[above right, font=\small] at (delta) {$\delta$};
    \filldraw[black] (delta) circle (1.5pt);

\node[below, align=center, font=\small, text width=8cm] at (axis description cs:0.5,-0.15)
        {};

    \end{axis}
    
\end{tikzpicture}
\caption{The ``phase space'' to determine the locations  of the extreme values of the helicity density ($0 < v < 1$). The signs in the brackets $(\bullet,\bullet)$ indicate the sign of ($f_+$, $f_-$)}\label{phase}
\end{figure}

\begin{figure}[ht]
    \centering
    \begin{tikzpicture}
     \begin{axis}[
        title={}, 
        title style={yshift=-5ex},
        xlabel={$v$}, 
        ylabel={$v_{\parallel}$}, 
        every axis x label/.style={at={(ticklabel* cs:1.00)},anchor=west, font=\small},
        every axis y label/.style={at={(ticklabel* cs:1.00)},anchor=south, font=\small},
        xmin=0, xmax=1.1,          
        ymin=0, ymax=1.2,      
        axis lines=left,         
        legend pos=outer north east, 
        domain=0:1,              
        samples=201,             
        smooth,                  
        legend cell align={left},
        xtick=\empty,
        ytick=\empty,
        clip mode=individual,
    ]
    \addplot[
        color=green,
        thick, 
        dashed,
    ] {x}; 
    \addlegendentry{$v_{\parallel} = v$};
    \addplot[
        name path=curveBlue,
        color=blue,
        thick,
        domain={1/3}:{1/sqrt(3)}, 
    ] {(1 - 3*x^2)/2};
    \addlegendentry{$v_{\parallel} = \frac{1-3v^2}{2}$}; 
    \addplot[
            color=purple,
            thick,
        ] {4*x^2/(1+3*x^2)};
        \addlegendentry{$v_{\parallel} = \frac{4v^2}{1+3v^2}$};
    \draw[dashed, gray] (axis cs:1,0) -- (axis cs:1,1);
    \node[below left, font=\small] at (axis cs:0,0) {$o$};
    \filldraw[black] (axis cs:0,0) circle (1.5pt);
\coordinate (alpha) at (axis cs:{1/3}, {1/3});
    \node[above right, font=\small] at (alpha) {$\alpha$};
    \filldraw[black] (alpha) circle (1.5pt);
 \coordinate (beta) at (axis cs:{1/sqrt(3)}, 0);
    \node[below right, font=\small] at (beta) {$\beta$};
    \filldraw[black] (beta) circle (1.5pt);
\coordinate (gamma) at (axis cs:1,1);
    \node[above left, font=\small] at (gamma) {$\gamma$};
    \filldraw[black] (gamma) circle (1.5pt);
\coordinate (delta) at (axis cs:1,0);
    \node[above right, font=\small] at (delta) {$\delta$};
    \filldraw[black] (delta) circle (1.5pt);

\node[below, align=center, font=\small, text width=8cm] at (axis description cs:0.5,-0.15)
        {};

    \end{axis}
\end{tikzpicture}\caption{The purple line is used to indicate the locations of the minimal values of the helicity density.}\label{posito}
\end{figure}

\be 
o=(0,0),\quad \alpha=(\frac{1}{3},\frac{1}{3}),\quad \beta=(\frac{1}{\sqrt{3}},0),\quad \gamma=(1,1).
\ee The dashed line is $v=1$ which can never be achieved. It will intersect  with $v_{\parallel}=0$ at the point \be \delta=(1,0).\ee

\begin{figure}[h!]
    \centering
    \includegraphics[width=0.9\textwidth]{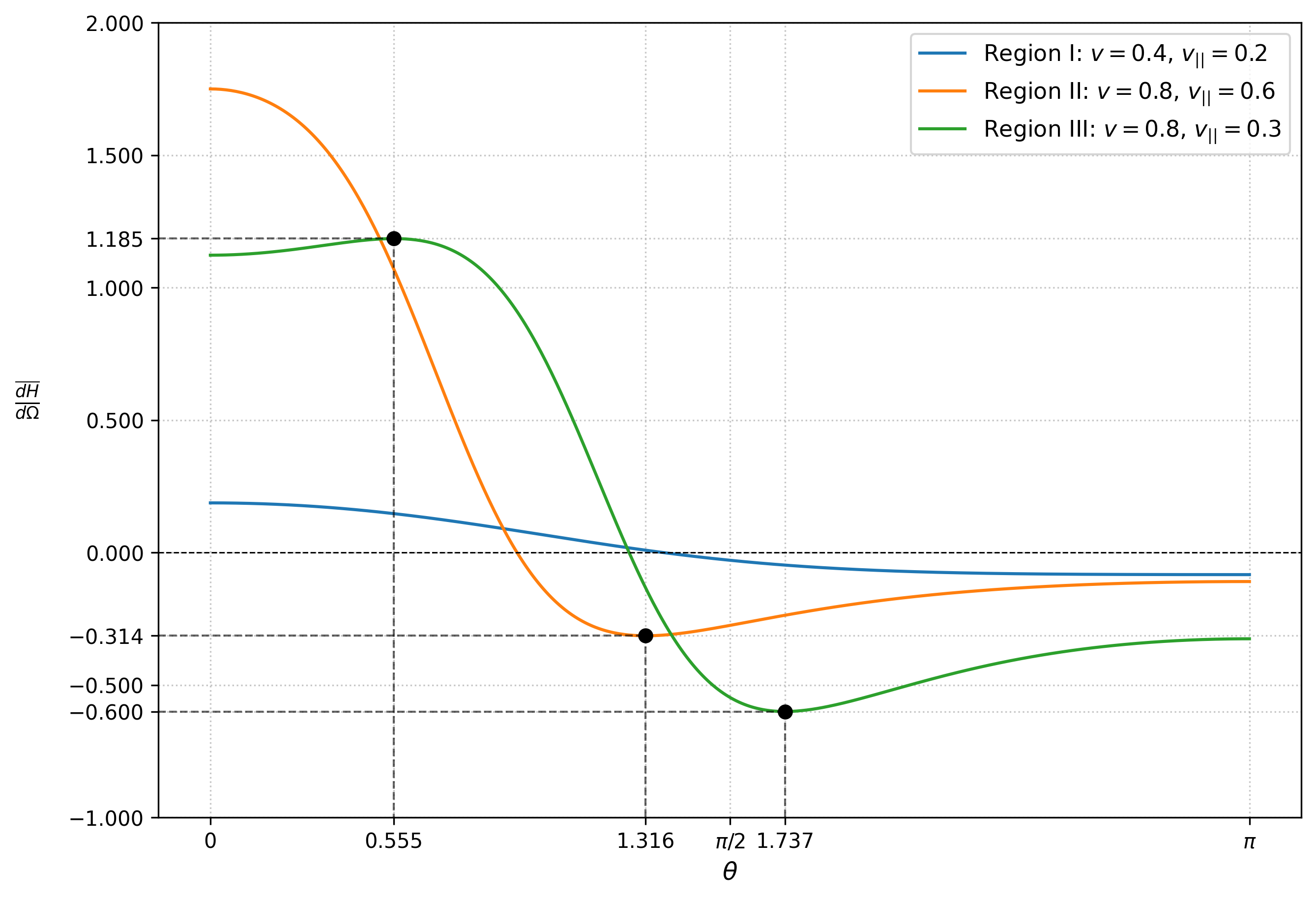}
    \caption{
        Angular distribution of the  rescaled helicity density $\frac{\overline {dH}}{d\Omega}$   for a helical motion. 
    }
    \label{fig:helical_motion_helicity}
\end{figure}

\begin{enumerate}
    \item Region I. This is bounded by the curved  triangle $o\alpha\beta$ with
    \be 
    f_-<0,\quad f_+<0.
    \ee In this case, both of $\lambda_+$ and $\lambda_-$ are not located between the interval $[-1,1]$. More precisely, we have $\lambda_-<-1,\ \lambda_+>1$. Therefore, there is no real solution for the angle $\theta$. The maximal value of the helicity density is located at the north pole and the minimal value at the south  pole
     \be 
    \frac{dH}{d\Omega}\Big|_{\text{max}}=\frac{q^2}{8\pi}\frac{v_{\perp}^2}{(1-v_{\parallel})^2},\quad  \frac{dH}{d\Omega}\Big|_{\text{min}}=-\frac{q^2}{8\pi}\frac{v_{\perp}^2}{(1+v_{\parallel})^2}.\label{maxH}
    \ee 
    \item Region II. This is bounded by the curved  triangle $\alpha\beta\gamma$ with 
    \be 
    f_->0,\quad f_+<0.
    \ee 
    In this case, we find $\lambda_+>1$ while $-1<\lambda_-<1$, the maximum  of the helicity density is still located at the north pole while the minimum  is located at 
    \be 
    \theta^{ II}_-=\arccos\lambda_-.\label{theta2}
    \ee The maximum is the same as the first equation of \eqref{maxH}  and the minimum  is 
    \be 
    \frac{dH}{d\Omega}\Big|_{\text{min}}=\frac{q^2v_{\perp}^2}{8\pi}\frac{\lambda_--v_{\parallel}}{\left((1-v_{\parallel}\lambda_-)^2-v_{\perp}^2(1-\lambda_-^2)\right)^{3/2}}.\label{mini}
    \ee 
    \item Region III. The signs of the functions $f_{\pm}$ are 
    \be 
    f_->0,\quad f_+>0.
    \ee From the purple curve of Figure \ref{posito}, we conclude that  $v_{\parallel}<\frac{4v^2}{1+3v^2}$ in this region. Therefore, we find  $0<\lambda_{\text{min}}<1$ and then $-1<\lambda_-<\lambda_+<1$, the maximum  of the helicity density is located at $\theta^{ III}_+$ while the minimum  is  located at $\theta^{ III}_-$ with
    \be 
    \theta^{ III}_{\pm}=\arccos\lambda_{\pm}.\label{theta3}
    \ee The maximum  is 
     \be 
    \frac{dH}{d\Omega}\Big|_{\text{max}}=\frac{q^2v_{\perp}^2}{8\pi}\frac{\lambda_+-v_{\parallel}}{\left((1-v_{\parallel}\lambda_+)^2-v_{\perp}^2(1-\lambda_+^2)\right)^{3/2}}\label{maxim}
    \ee and the minimum is still given by \eqref{mini}. One can prove that $\lambda_-< v_{\parallel}< \lambda_+$. As a consequence, the minimum(maximum)  of the helicity density is always negative(positive).
\end{enumerate}
{ Figure \ref{fig:helical_motion_helicity} shows the angular distribution of the rescaled helicity density for a charged particle in helical motion
\be 
\frac{\overline{dH}}{d\Omega}=\frac{v^2_\perp(\cos\theta-v_{\parallel})}{\left((1-v_{\parallel}\cos\theta)^2-v_{\perp}^2\sin^2\theta\right)^{3/2}}.
\ee  The helicity density depends only on the polar angle $\theta$.
        The plot shows three distinct cases, each corresponding to one of the three regions analyzed in the $v$-$v_{||}$ phase space: Region I ($v=0.4, v_{||}=0.2$), Region II ($v=0.8, v_{||}=0.6$), and Region III ($v=0.8, v_{||}=0.3$). 
        The black dots mark the locations of the extrema. }

\subsubsection{Soft Bremsstrahlung}
\begin{figure}[h!]
    \centering
    \includegraphics[width=\textwidth]{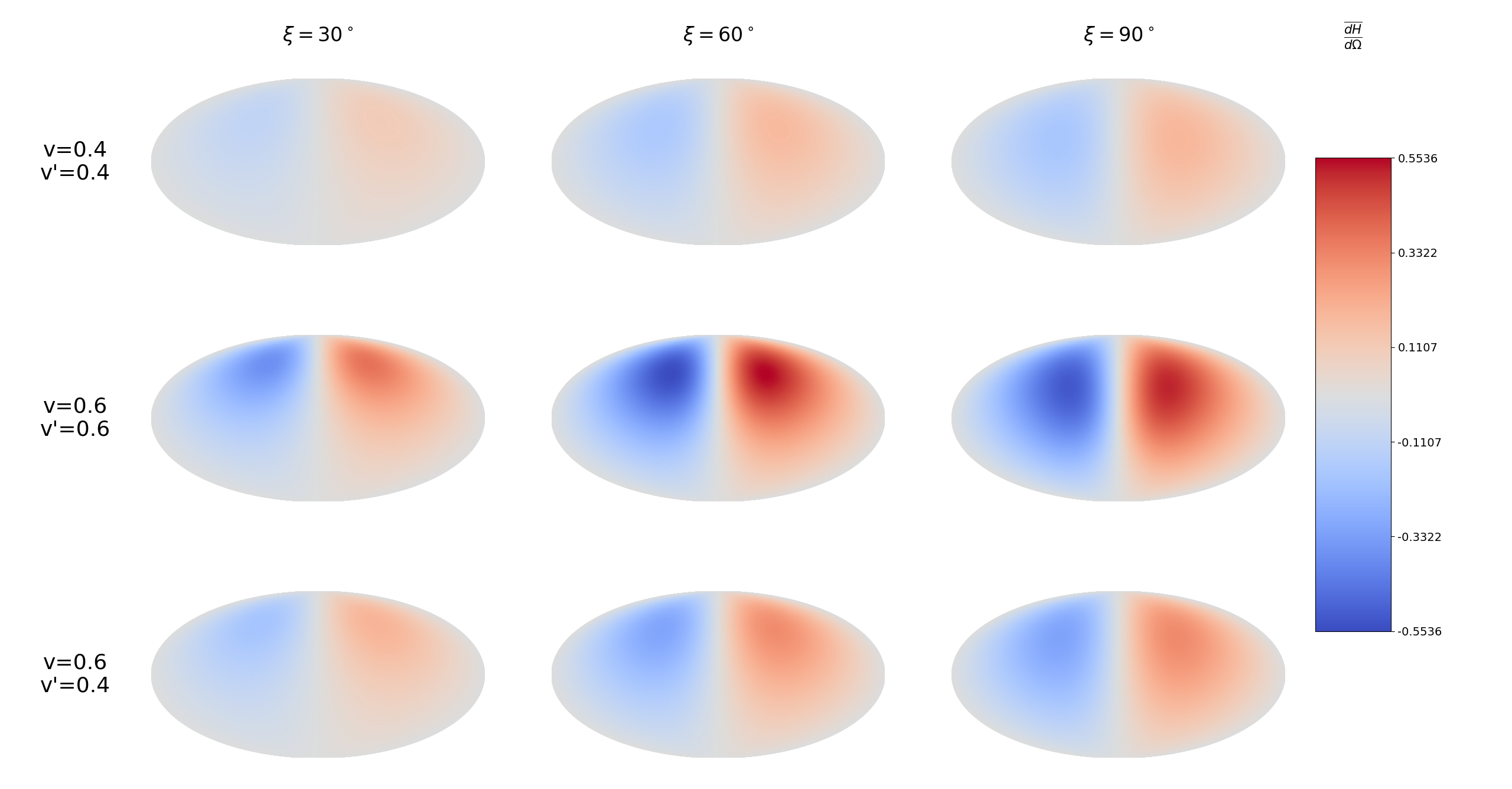}
    \caption{
        The rescaled helicity density $\frac{\overline{dH}}{d\Omega} $ for the soft bremsstrahlung. 
    }
    \label{fig:helicity_distribution}
\end{figure}

In this situation, a charge moving with velocity $\bm v$ receives a kick at time $t=0$, causing its velocity change to $\bm v'$. The four-current can be idealized as 
\be 
j^\mu(t,\bm x)=q(1,\bm v)\delta^{(3)}(\bm x-\bm v t)\theta(-t)+q(1,\bm v')\delta^{(3)}(\bm x-\bm v' t)\theta(t).
\ee The vector field at null infinity is 
\bea 
A_i(u,\Omega)&=&\frac{1}{4\pi}\int d^3\bm x' j_i(u+\bm n\cdot\bm x',\bm x')=\frac{1}{4\pi}[\frac{q v_i}{1-\bm n\cdot\bm v}\theta(-u)+\frac{qv_i'}{1-\bm n\cdot\bm v'}\theta(u)].
\eea In frequency space, we have 
\bea 
A_i(\omega,\Omega)=\int_{-\infty}^\infty du A_i(u,\Omega)e^{-i\omega u}=\frac{1}{4\pi}[\frac{iqv_i}{(1-\bm n\cdot\bm v)(\omega+i\epsilon)}-\frac{iqv_i'}{(1-\bm n\cdot\bm v')(\omega-i\epsilon)}].
\eea Utilizing the formula \eqref{domegaE}, we obtain the energy flux density in frequency space
\bea 
\frac{dE}{d\omega d\Omega}=\frac{q^2}{16\pi^3}[\frac{2(1-\bm v\cdot\bm v')}{(1-\bm n\cdot\bm v)(1-\bm n\cdot\bm v')}-\frac{1-\bm v^2}{(1-\bm n\cdot\bm v)^2}-\frac{1-\bm v'^2}{(1-\bm n\cdot\bm v')^2}].\label{energyfluxtwo}
\eea The angular distribution in the square brackets is exactly the one in \cite{1995iqft.book.....P}. Similarly, we find the helicity flux density in frequency space 
\bea 
\frac{dH}{d\omega d\Omega}&=&-\frac{iq^2}{32\pi^3}\omega n_i \epsilon_{ijk}[(\frac{v_j}{(1-\bm n\cdot\bm v)(\omega+i\epsilon)}-\frac{v_j'}{(1-\bm n\cdot\bm v')(\omega-i\epsilon)})\nn
\\
&&\times(\frac{v_k}{(1-\bm n\cdot\bm v)(\omega-i\epsilon)}-\frac{v_k'}{(1-\bm n\cdot\bm v')(\omega+i\epsilon)})-(j\leftrightarrow k)]\nn\\&=&\frac{-iq^2}{8\pi^3} \frac{n_i\epsilon_{ijk}v_j v_k'}{(1-\bm n\cdot\bm v)(1-\bm n\cdot\bm v')}\omega [\frac{1}{(\omega-i\epsilon)^2}-\frac{1}{(\omega+i\epsilon)^2}].
\eea Note that the $i\epsilon$ is important in the expression. One can use the formula \footnote{Here `PV' means principal value.}
\bea 
\frac{1}{\omega\pm i\epsilon}=\text{PV}\left(\frac{1}{\omega}\right)\mp \pi i \delta(\omega)
\eea to obtain 
\bea 
\frac{dH}{d\omega d\Omega}&=&-\frac{q^2}{4\pi^2}\frac{\bm n\cdot(\bm v\times\bm v')}{(1-\bm n\cdot\bm v)(1-\bm n\cdot\bm v')}\omega\delta'(\omega)=\frac{q^2}{4\pi^2}\frac{\bm n\cdot(\bm v\times\bm v')}{(1-\bm n\cdot\bm v)(1-\bm n\cdot\bm v')}\delta(\omega).
\eea We used the notation $\delta'(\omega)=\frac{d}{d\omega}\delta(\omega)$. In the last step, we have used the identity for the Dirac delta function 
\be 
\omega \delta'(\omega)=-\delta(\omega)
\ee for well behaved test functions. There is a peak at $\omega=0$. After integrating over the frequency space, we find the helicity density 
\be 
\frac{dH}{d\Omega}=\frac{q^2}{4\pi^2}\frac{\bm n\cdot(\bm v\times\bm v')}{(1-\bm n\cdot\bm v)(1-\bm n\cdot\bm v')}.
\label{88}
\ee The helicity density vanishes along directions that are perpendicular to the plane generated by $\bm v$ and $\bm v'$. When $\bm v$ and $\bm v'$ are collinear, the helicity density is always zero in all directions. After some efforts, one can prove that the radiative helicity is zero \footnote{This can be proved as follows. The integral on the sphere 
\be 
\int d\Omega \frac{\bm n}{(1-\bm n\cdot\bm v)(1-\bm n\cdot \bm v')}
\ee can be fixed to two terms by the index structure 
\bea 
\int d\Omega \frac{\bm n}{(1-\bm n\cdot\bm v)(1-\bm n\cdot\bm v')}=f_1 \bm v+f_2 \bm v'.
\eea The right hand side is perpendicular to $\bm v\times\bm v'$ and thus the integral in \eqref{radiativesoft} {  vanishes}. One can find more complete analysis for such kind of integrals in Appendix \ref{int}.  }
\be 
H=\int d\Omega \frac{dH}{d\Omega}=0.\label{radiativesoft}
\ee 
{  Figure \ref{fig:helicity_distribution} shows the angular distribution of the rescaled helicity density for soft bremsstrahlung
\bea 
\frac{\overline{dH}}{d\Omega}=\frac{\bm n\cdot(\bm v\times\bm v')}{(1-\bm n\cdot\bm v)(1-\bm n\cdot\bm v')}.
\eea  
The direction of $\bm v$ is fixed to align with the $x$-axis. 
The other velocity $\bm v'$ in spherical coordinates is
\be 
\bm v'=(v',\theta'=\frac{\pi}{2}-\xi,\phi'=0).
\ee 
We use the Mollweide projection to represent the celestial sphere as an ellipse. This technique is a standard tool in cosmic microwave background (CMB) observations.
The top row displays results for  speeds $v = v' = 0.4$, the second row for speeds $v = v' = 0.6$, and the bottom row for $v = 0.6$, $v' = 0.4$. In all cases, the angle $\xi$ between the two velocities varies from $30^\circ$ to $90^\circ$.

}

\subsubsection{Oscillating electric and magnetic dipoles}\label{Oscelemagdi}
In previous toy models, the radiative helicity is always zero. In this section, we will introduce a model whose helicity flux is non-vanishing. 
In this model, the system consists of an electric dipole of moment $p_0$ pointing in the $z$-direction, oscillating with frequency $\omega$ and a  magnetic dipole of moment $m_0$ is located in $x$-$y$ plane, oscillating with frequency $\omega'$. The vector potential is the superposition of the ones generated by the electric and magnetic dipoles
\bea 
A_i(u,\Omega)=-\frac{1}{4\pi}p_0\omega\cos\omega u\bm e_z-\frac{1}{4\pi}m_0\omega'\sin\omega'u \sin\theta(\cos\phi\bm e_y-\sin\phi\bm e_x).
\eea Then the helicity flux density is 
\be 
\frac{dH}{du d\Omega}=-\frac{p_0m_0\omega\omega'\sin^2\theta}{16\pi^2}(\omega\sin\omega u\sin\omega'u+\omega'\cos\omega u\cos\omega'u)
\ee which is independent of the retarded time for $\omega'=\omega$
\be 
\frac{dH}{du d\Omega}=-\frac{p_0m_0\omega^3\sin^2\theta}{16\pi^2}.
\ee Note that the helicity flux density has an angle-distribution over the sphere and the corresponding helicity flux is non-zero 
\be 
\frac{dH}{du}=-\frac{p_0m_0\omega^3}{6\pi}.
\ee 

\section{Multipole expansion of the EM helicity flux}\label{mul}
In this section, we will discuss the multipole expansion of the EM helicity flux density. We will assume the source is located at a finite region such that we may expand the current order by order 
\bea 
A_i(u,\Omega)=\frac{1}{4\pi}\int_V d\bm x' \sum_{n=0}^\infty \frac{1}{n!}j_i^{(n)}(u,\bm x')(\bm n\cdot\bm x')^n\label{expansion}
\eea where $j_i^{(n)}=\frac{d^n}{du^n}j_i$. One can turn it into two sets of multipoles by reorganising terms and gauge transformation
\bea 
A_i(u,\Omega)=\frac{1}{4\pi}\sum_{\ell=1}^\infty \frac{n^{\text{STF}}_{j_1\cdots j_{\ell-1}}}{\ell!}\left(\frac{d^\ell}{du^{\ell}} I_{ij_1\cdots j_{\ell-1}}-\epsilon_{ikj_\ell}n_{k}\frac{d^\ell}{du^\ell}J_{j_1\cdots j_\ell}\right)\label{Aiex}
\eea 
where $I_{i_1\cdots i_\ell}$ and $J_{i_1\cdots i_\ell}$ are symmetric and traceless electric and magnetic multipoles, respectively. The tensor $n_{j_1\cdots j_\ell}^{\text{STF}}$ is the symmetric and traceless tensor by subtracting the traces from $n_{j_1}n_{j_2}\cdots n_{j_\ell}$. {The multipole expansion \eqref{Aiex} can be compared with eqn.(4.16) of \cite{Damour:1990gj} and eqn.(54)-(55) of \cite{Ross_2012}. 
The explicit forms of $I_{i_1\cdots i_{\ell}}$ and $J_{i_1\cdots i_\ell}$ can be found, for example, in eqn.(47)-(48) of  \cite{Ross_2012}\footnote{  We have redefined the normalization of the magnetic multipole moment $J_{i_1\cdots i_\ell}$ compared with \cite{Ross_2012}, resulting in a more concise and symmetrical expression. More precisely,   \begin{align}
I_{i(\ell)}\equiv I_{i_1\cdots i_{\ell}}&=\sum_{p=0}^{\infty}{\frac{\left( 2\ell +1 \right) !!}{\left( 2p \right) !!\left( 2\ell +2p+1 \right) !!}\left\{ \frac{\ell}{\ell +2p} \right. \left[ \int{d^3\boldsymbol{x}\partial _{u}^{2p}\rho r^{2p}x_{{i_1}}}\cdots x_{i_{\ell}} \right] ^{\mathrm{STF}}}\nn
\\
{}&+\frac{\ell}{\left( \ell +1 \right) \left( \ell +2p+2 \right)}\left. \left[ \int{d^3\boldsymbol{x}r^{2p}\partial _{u}^{2p+1}\left( j_{i_{\ell}}x_{i_1}\cdots x_{i_{\ell -1}}r^2-\boldsymbol{j}\cdot \boldsymbol{x}\,x_{i_1}\cdots x_{i_{\ell}} \right)} \right] ^{\mathrm{STF}} \right\} ,\label{elemulti}
\\
J_{i(\ell)}\equiv J_{i_1\cdots i_{\ell}}&=\frac{\ell}{\ell+1}\sum_{p=0}^{\infty}{\frac{\left( 2\ell +1 \right) !!}{\left( 2p \right) !!\left( 2\ell +2p+1 \right) !!}\left[ \int{d^3\boldsymbol{x}\partial _{u}^{2p}\left( \boldsymbol{x}\times \boldsymbol{j} \right) _{i_{\ell}}x_{i_1}}\cdots x_{i_{\ell -1}}r^{2p} \right] ^{\mathrm{STF}}},\label{magmulti}
\end{align} 
where the shorthand notation $i(\ell)=i_i\cdots i_\ell$ denotes symmetric trace free (STF) indices. }. 
In general, the electric multipole moments $I_{i(\ell)}$ receive contributions not only from the electric charge density but also from the electric current density. In contrast, the magnetic multipole moments are generated solely by the electric current density. Furthermore, we note that the expression $I_{i(\ell)}$ is not unique due to the conservation law $\partial_\mu j^\mu=0$.  There is  an equivalent approach to defining these multipoles via Debye potentials and spherical harmonics \cite{PhysRevD.15.2156}, since both of them are irreducible representations of $SO(3)$ in three-dimensional space. For slowly varying sources,  the $p=0$ terms provide the leading order contributions, and the electric and magnetic multipoles can be approximated by 
\begin{align}
     I_{i(\ell)}=& \left[\int d^3\bm x \rho x_{i_1}\cdots x_{i_\ell}\right]^{\text{STF}},\label{Iil}\\
     J_{i(\ell)}=&\frac{\ell}{\ell+1} \left[\int d^3\bm x (\bm x\times\bm j)_{i_\ell}x_{i_1}\cdots x_{i_{\ell-1}}\right]^{\text{STF}}.\label{Jil}
\end{align}
}

To check the previous formulae, we consider the leading term in the expansion \eqref{expansion}
\bea 
A_i(u,\Omega)&\approx &\frac{1}{4\pi}\int_V d^3\bm x' j_i(u,\bm x')=\frac{1}{4\pi}\int_V d\bm x' [\partial'_{j}(x'_i j_j(u,\bm x'))-x'_i \partial'_j j_j(u,\bm x')]
\eea where the first term in the brackets vanishes via integration by parts and the second term can be transformed to 
\bea 
A_i(u,\Omega)&\approx & \frac{1}{4\pi}\int_V d^3\bm x'(x'_i \partial_u \rho(u,\bm x'))=\frac{1}{4\pi}\frac{d}{du}p_i(u)\label{Aip}
\eea where $p_i(t)$ is defined as the electric dipole at time $t$
\be 
p_i(t)=\int_V d^3\bm x \rho(t,\bm x)x_i.
\ee
{  On the other hand, the $\ell=1$ term of \eqref{Iil} reads 
\begin{align}
I_i&=\left[ \int{d^3\boldsymbol{x}}\rho x_i \right] ^{\mathrm{STF}}= p_i.
\end{align}
It is nothing but the definition of the  electric dipole moment.} Note that \eqref{Aip} indicates that $A_i(u,\Omega)$ may be explained as an extended (time derivative of) electric dipole since its leading order is proportional to $\dot p_i$. Therefore, the electric dipole contributes to the 
energy/helicity flux density as follows
\begin{align}\label{electricdipole}
\frac{dE}{du d\Omega}\Big|_{\text{electric dipole}}&=\frac{1}{16\pi^2}[\ddot {\bm p}^2-(\bm n\cdot\ddot p)^2],\\
\frac{dH}{du d\Omega}\Big|_{\text{electric dipole}}&=\frac{1}{16\pi^2}\bm n\cdot(\dot{\bm p}\times\ddot{\bm p}).
\end{align}
By integrating over the sphere, the energy and helicity flux are
\bea 
\frac{dE}{du}&=&\frac{\ddot{\bm p}^2}{6\pi},\\
\frac{dH}{du}&=&0.
\eea 
\begin{enumerate}
    \item For an electric dipole that is oscillating in a fixed direction with frequency $\omega$ 
\be 
\bm p=p_0\cos\omega u\bm e_z,
\ee one may compute the energy flux density 
\be 
\frac{dE}{du d\Omega}=\frac{1}{16\pi^2}p_0^2\omega^4\sin^2\theta \cos^2\omega u
\ee and the average energy flux in a period 
\be 
\langle \frac{dE}{du}\rangle=\frac{p_0^2\omega^4}{12\pi}
\ee which is consistent with \cite{jackson1999classical}. Note that $\dot {\bm p}$ is collinear with $\ddot{\bm p}$, therefore, the helicity flux density is always zero.
\item For an electric dipole that is oscillating in a plane with frequency $\omega$
\be 
\bm p=p_0\cos\omega u\bm e_x+p_0\sin\omega u\bm e_y,
\ee both of the energy flux density and helicity flux density are non-vanishing 
\bea 
\frac{dE}{du d\Omega}&=&\frac{p_0^2\omega^4}{16\pi^2}\left(1-n_2^2\cos^2\omega u-n_3^2\sin^2\omega u\right),\\
\frac{dH}{du d\Omega}&=&\frac{p_0^2\omega^3}{16\pi^2}\cos\theta.
\eea 
\end{enumerate}

Up to the subleading order, a straightforward calculation leads to 
\be 
A_i(u,\Omega)\approx \frac{1}{4\pi}\int d\bm x' (j_i(u,\bm x')+\dot{j}_i(u,\bm x')\bm n\cdot\bm x')=\frac{1}{4\pi}[\dot p_i(u)+\dot\Lambda_{ij}n_j]
\ee where the tensor $\Lambda_{ij}$ is decomposed as
\be 
\Lambda_{ij}=\frac{1}{2}\frac{d}{du} D_{ij}(u)+\frac{1}{6}\delta_{ij}S-\epsilon_{ijk}m_k(u)
\ee where $D_{ij}$ is the electric quadrupole moment 
\be 
D_{ij}(t)=\int d^3\bm x \rho(t,\bm x)(x_i x_j-\frac{1}{3}\delta_{ij}\bm x^2)
\ee and $m_i$ is the magnetic dipole\cite{jackson1999classical} 
\be 
m_i(t)={\frac{1}{2}}\int d^3\bm x \epsilon_{ijk}x_j j_k(t,\bm x).
\ee The second term $S$ is a trace which can be gauged away. In other words, this term will not  contribute to the transverse mode due to the identity 
\be 
Y^i_A n_i=0.
\ee Therefore, we can ignore this term and write $A_i$ as follows
\be 
A_i(u,\Omega)=\frac{1}{4\pi}[\dot p_i(u)+\frac{1}{2}\ddot{D}_{ij}n_j-\epsilon_{ijk}n_j \dot m_k(u)+\cdots]\label{Ai}
\ee where `$\cdots$'  denotes higher-order multipole contributions.{  Similarly, taking  
the $\ell=1$ term of \eqref{Jil}
\bea
J_i=\left[\int d^3\bm x\left(\bm x\times \bm j\right)_i\right]^{\text{STF}}= m_i,
\eea
we obtain the definition of magnetic dipole moment straightforwardly. Meanwhile, the $\ell=2$ term of  \eqref{Iil}
\bea 
I_{ij}=\left[\int d^3\bm x\rho x_{i}x_{j}\right]^{\text{STF}}= D_{ij},
\eea 
is exactly the definition of the electric quadrupole moment.
}The energy and helicity flux density can be obtained by substituting \eqref{Ai} into \eqref{energyflux1} and \eqref{helicityflux1}, respectively. We will only present the contribution of the magnetic dipole moment
\begin{align} \label{magneticdipole}
\frac{dE}{du d\Omega}\Big|_{\text{magnetic dipole}}&=\frac{1}{16\pi^2}(\ddot{\bm m}^2-(\ddot{\bm m}\cdot\bm n)^2),\\
\frac{dH}{du d\Omega}\Big|_{\text{magnetic dipole}}&=\frac{1}{16\pi^2}\bm n\cdot(\dot{\bm m}\times\ddot{\bm m}).\label{hedipole}
\end{align} We note that the structure is completely the same as the contribution of the electric dipole moment in \eqref{electricdipole}. This formula will be useful when we consider the EM helicity flux density from pulsar systems. The complete expressions for the energy and helicity flux density, including all multipole contributions, are given by
\begin{align}
T(u,\Omega)&=\frac{P_{ip}}{16\pi^2}\left(\sum_{\ell=1}^\infty \frac{n^{\text{STF}}_{j(\ell-1)}}{\ell!}(I^{(\ell+1)}_{ij(\ell-1)}-{}\epsilon_{ikj_{\ell}}n_k J^{(\ell+1)}_{j(\ell)})\right)\nonumber\\
&\times\left(\sum_{\ell'=1}^\infty \frac{n^{\text{STF}}_{q(\ell'-1)}}{\ell'!}(I^{(\ell'+1)}_{pq(\ell'-1)}-\epsilon_{pmq_{\ell'}}n_m J^{(\ell'+1)}_{q(\ell')})\right),\\
O(u,\Omega)&=\frac{\epsilon_{ipp'}n_{p'}}{16\pi^2}\left(\sum_{\ell=1}^\infty \frac{n^{\text{STF}}_{j(\ell-1)}}{\ell!}(I^{(\ell)}_{ij(\ell-1)}-\epsilon_{ikj_{\ell}}n_k J^{(\ell)}_{j(\ell)})\right)\nonumber\\
&\times\left(\sum_{\ell'=1}^\infty \frac{n^{\text{STF}}_{q(\ell'-1)}}{\ell'!}(I^{(\ell'+1)}_{pq(\ell'-1)}-\epsilon_{pmq_{\ell'}}n_m J^{(\ell'+1)}_{q(\ell')})\right)\label{multiheflden}
\end{align} where 
\be 
I^{(p)}_{i(\ell)}=\frac{d^p}{du^p}I_{i(\ell)},\quad J^{(p)}_{i(\ell)}=\frac{d^p}{du^p}J_{i(\ell)},
\ee


Integrating over the sphere, the energy and helicity flux are
\begin{align} 
\frac{dE}{du}&=\frac{1}{4\pi}\sum_{\ell=1}^\infty \frac{m_{\ell-1}-m_\ell}{\ell!^2}\left(|I_{i(\ell)}^{(\ell+1)}|^2+|J_{i(\ell)}^{(\ell+1)}|^2\right),\label{enerhelicity}\\
\frac{dH}{du}&=-\frac{1}{4\pi}\sum_{\ell=1}^\infty \frac{m_{\ell-1}-m_{\ell}}{\ell!^2}\times \left(I^{(\ell)}_{i(\ell)}J^{(\ell+1)}_{i(\ell)}-J^{(\ell)}_{i(\ell)}I^{(\ell+1)}_{i(\ell)}\right)\label{helicityf}
\end{align} where 
\bea 
m_\ell=\frac{\ell!}{\left(2\ell+1\right)!!}.
\eea

The formula \eqref{enerhelicity} is the EM radiant power whose gravitational analog has been derived in \cite{Thorne:1980ru}. {  It contains only square terms of electric and magnetic multipoles, such as $|I^{(\ell+1)}_{i(\ell)}|^2$ and $|J^{(\ell+1)}_{i(\ell)}|^2$.} Switching to the frequency space, we checked that it is consistent with the one in \cite{Ross_2012}.

The second formula \eqref{helicityf} has not been derived in the literature so far. Note that similar formulae for the  gravitational helicity flux can be found in \cite{Long:2024yvj}. 
{  As shown in \eqref{helicityf}, the total helicity flux lacks pure electric or magnetic multipoles of the form $|I^{(\ell+1)}_{i(\ell)}|^2$ or $|J^{(\ell+1)}_{i(\ell)}|^2$ unlike the energy flux. Instead, it is entirely composed of cross terms between electric and magnetic multipoles. This suggests that the net radiative helicity flux requires the radiation source to simultaneously excite specific electric and magnetic multipoles. The oscillating electric and magnetic dipole model discussed in Subsection \ref{Oscelemagdi} provides a direct example of this mechanism: only when both dipoles coexist and oscillate in non-parallel orientations can the system radiate net helicity into space. }

\section{Discussion}\label{pul}
We have derived several  formulae for EM radiative helicity and its angular distribution. Now we will discuss the potential applications of the previous results.

\paragraph{Pulsar systems.} Pulsar {systems} and their applications in astronomy have been reviewed in the book \cite{2006puas.book.....L}. Pulsars are rapidly spinning neutron stars which are extremely dense and composed almost entirely of neutrons. More importantly, they emit periodic pulses  of EM 
radiation which are ideal laboratories to check the EM helicity flux density. All pulsars lose energy, either to magnetic dipole radiation or to charged particle winds. In this work, we will consider a conventional model that  a pulsar loses its energy due to magnetic dipole radiation \cite{deutsch1955electromagnetic,pacini1968rotating}. In this model, the rotation axis of the pulsar is along $z$-direction and magnetic axis is along a rotating magnetic dipole moment $\bm m(t)$ inclined at an angle $\chi$
\bea 
\bm m(t)=m_0\sin\chi \cos(\omega t)\bm e_x+m_0\sin\chi \sin(\omega t)\bm e_y+m_0\cos\chi \bm e_z
\eea where $m_0$ is the magnitude of the dipole and $\omega$ is the frequency which is related to the period $T$ through the equation 
\be 
\omega=\frac{2\pi}{T}.
\ee 

Substituting into \eqref{magneticdipole} and \eqref{hedipole}, the energy and helicity flux density are \bea 
T(u,\Omega)=\frac{dE}{du d\Omega}&=&\frac{1}{16\pi^2}m_0^2\omega^4\sin^2\chi[\sin^2(\omega u-\phi)+\cos^2\theta \cos^2(\omega u-\phi)],\\
O(u,\Omega)=\frac{dH}{du d\Omega}&=&\frac{1}{16\pi^2}m_0^2\omega^3\sin^2\chi \cos\theta.\label{hd}
\eea 
In particular, these equations become consistent with those derived in \cite{deutsch1955electromagnetic} for the far-field regime, provided the magnetic dipole  $m_0$ satisfies the relation:
\bea
m_0=\frac{1}{2} B_0r_0^3 
\eea where $r_0$ is the radius of the pulsar and $B_0$ is  magnetic field strength on the surface of the pulsar.

The average energy flux density is 
\be 
\langle T(u,\Omega)\rangle=\langle \frac{dE}{du d\Omega}\rangle=\frac{1}{32\pi^2}m_0^2\omega^4\sin^2\chi(1+\cos^2\theta).\label{avT}
\ee Therefore, the average energy flux is 
\be 
P_{\text{rad}}=\langle \frac{dE}{du}\rangle =\frac{1}{6\pi}m_0^2\omega^4\sin^2\chi
\ee which agrees with radiant power contributed by a magnetic dipole. Interestingly, the helicity flux density is also proportional to the square of the perpendicular magnetic dipole $m_{\perp}=m_0\sin\chi$. Therefore, we obtain the ratio between the average helicity and energy flux density 
\be 
\frac{\langle O(u,\Omega)\rangle}{\langle T(u,\Omega)\rangle}=\frac{2}{\omega}\frac{ \cos\theta}{1+\cos^2\theta}=\frac{T}{\pi}\frac{ \cos\theta}{1+\cos^2\theta}.
\label{magnitude}
\ee The ratio is independent of the inclination angle and proportional to the  pulsar rotation period. The maximal/minimum ratio is located at the north/south pole. 
An interesting inequality is followed
\be 
T\ge 2\pi \Big|\frac{\langle O(u,\Omega)\rangle}{\langle T(u,\Omega)\rangle}\Big|.
\ee 
We will use this formula to estimate the magnitude of the helicity flux density. The characteristic magnitude is defined as
\be 
E_{c}=P_{\text{rad}}T\label{chara}
\ee where $P_{\text{rad}}$ is the radiant power and $T$ is the period. In practice, the pulsar's period will increase with time.  For Crab Pulsar (PSR B0531+21), a remnant 
 of the supernova SN 1054,  they are approximately  \cite{ljg} 
 \be 
 T=33\text{ms},\quad \dot T=4.2\times 10^{-13}.
 \ee The energy loss is dominated by the spin down luminosity, which is related to the rate of loss of the rotational kinematic energy \cite{lorimer2012handbook}
 \be
 P_{\text{rad}}=4\pi^2I\frac{\dot{T}}{T^3},
 \ee where the momenta of inertia is 
 \be 
 I=10^{45}\text{g}\cdot \text{cm}^2.
 \ee 
 Therefore, the characteristic magnitude of the helicity flux density is 
 \be 
 E_c\approx 1.5\times 10^{30}\text{J}=1.5\times 10^{30}\text{kg}\cdot  \text{m}^2/\text{s}^2.
 \ee 
 The distance between the Crab Pulsar and the solar system is approximately (obtained from the ATNF Pulsar Catalogue \cite{Manchester2005_ATNF}\footnote{\href{https://www.atnf.csiro.au/research/pulsar/psrcat/}{https://www.atnf.csiro.au/research/pulsar/psrcat/}.})
 \be 
 d=2\text{kpc}\approx6500 \text{ly}=6.15\times 10^{19}\text{m}.
 \ee Therefore, the magnitude of the helicity flux density on Earth is approximately 
 \be 
 \frac{E_c}{4\pi d^2}\approx 3\times 10^{-11}\text{J}/\text{m}^2.
 \ee 
 

Radiation emission from pulsars is a complicated problem which raises the issue of the structure of the neutron star magnetosphere. The problem is firstly tackled by 
\cite{1969ApJ...157..869G,1982RvMP...54....1M} and it is expected that the force-free MHD equations  provide a very good approximation in the magnetosphere \cite{1973ApJ...180L.133M,Blandford:1977ds}. Unfortunately, the force-free MHD equations are highly non-linear and one should use numerical simulation \cite{1999ApJ...511..351C} to study this problem. The numerical simulation leads to a more accurate radiant power \cite{2006ApJ...648L..51S}
\be 
P_{\text{rad}}\propto (1+\sin^2\chi)
\ee 
contrasts with \eqref{avT}.
Therefore, one should expect that the form of the helicity flux density \eqref{hd} may be modified after taking into account the electron-positron plasma in the magnetosphere.  Moreover, a related question is the nature of the spin-down. The rotation kinematic energy of the pulsar transfers to the radiative emission energy and causes the variation of the inclination angle and the rotation period \cite{1970ApL.....5...21M,2014MNRAS.441.1879P}. Therefore, the characteristic magnitude \eqref{chara} will evolve over time.

\paragraph{Solar and stellar systems.} EM helicity flux density is a novel quantity that is closely related to magnetic helicity.  In the last 70 years, the magnetic helicity has been applied to various systems, including solar flares and coronal mass ejections \cite{2002A&ARv..10..313P},  hydrodynamics and plasmas \cite{1969JFM....35..117M}, etc. Interestingly, even though it is obviously gauge dependent, the magnetic helicity density (per unit surface) has been studied in \cite{Welsch_2003,2005A&A...439.1191P,Lund_2020}. Numerous authors have been attempts to measure the 
 magnetic helicity density in the solar atmosphere, see reviews  by \cite{DEMOULIN20071674,DEMOULIN20091013}.  The EM helicity flux density distinguishes from it and is suitable for dynamical systems with magnetic field evolution.  As established in Section \ref{optical}, the EM helicity flux is, although not equivalent,  closely related to the optical helicity flux. The latter is observationally determined via the Stokes V parameter for monochromatic plane wave \cite{bliokh2015transverse}. On the other hand, Stokes parameter V is also used to reconstruct the topology of the magnetic field through stellar surface \cite{johnstone2010modelling, arzoumanian2011contribution, lang2014modelling, vidotto2016magnetic} which can be used to calculate the magnetic helicity density of surface magnetic field. Therefore, high-precision polarization measurement instruments \cite{Rimmele_2020,lites2013hinode,bellot2019quiet} provide probabilities of measuring the EM helicity flux density. \iffalse It is possible to make an order-of-magnitude estimate using (\ref{magnitude}), although there is currently no evidence to support describing the solar system's radiation with a simple rotating magnetic dipole model. Using the data provided by \cite{kopp2011new, Hathaway2015}, we can get the order-of-magnitude estimate $O\approx 4.7\times 10^{11} \text{J}/\text{m}^2$?}\fi
We believe that the EM helicity flux density will provide rich information besides magnetic helicity. The magnetic helicity is widely used in the process of magnetic reconnection \cite{2000mare.book.....P} which is common in evaluating solar and stellar activities, it would be interesting to explore this topic from the perspective of the EM helicity flux density
.

\vspace{10pt}
{\noindent \bf Acknowledgments.} 
The work of J.L. was supported by NSFC Grant No. 12005069.

\appendix
\section{Magnetic helicity, radiative helicity and electromagnetic Hopfion}\label{magsupp}
In this appendix, we will discuss more on the magnetic helicity and radiative helicity in time-dependent  Hopfions. In \cite{ranada1990knotted}, the authors obtained a time-dependent Hopfion in Maxwell theory. The 
explicit solution is \bea 
\bm e=\sqrt{\lambda}(p \bm H_1-q\bm H_2),\quad \bm b=\sqrt{\lambda}(q \bm H_1+p\bm H_2),
\eea where
\bea 
p=t(6a^2-8t^2),\quad q=a(a^2-12t^2),\quad a=1+x^2+y^2+z^2-t^2
\eea and the vectors $\bm H_1, \bm H_2$ are
\bea 
\bm H_1&=&\frac{4}{\pi d^3}(2(t+y-x z),-2(x+(t+y)z),-1+x^2+(t+y)^2+z^2),\\
\bm H_2&=&\frac{4}{\pi d^3}((t+y)^2-x^2+z^2-1,2 (z-x (t+y)),-2 (t+x z+y))
\eea with
\bea 
d&=a^2+4t^2.
\eea By expanding the solution near $\mathcal{I}^+$, we find the vector field  \begin{align}
A_1&=\frac{\sqrt{\lambda } \left(n_y \left(u^2-1\right) \left(n_x^2+n_z^2+1\right)+2 n_z \left(n_z \left(u^2-1\right)-2 n_x u\right)+n_y^3 \left(u^2-1\right)+2 n_y^2 \left(u^2-1\right)\right)}{4 \pi  \left(u^2+1\right)^2},\\
A_2&=\frac{\sqrt{\lambda } \left(n_x^2+(n_y+1)^2+n_z^2\right) \left(-n_x u^2+n_x-2 n_z u\right)}{4 \pi  \left(u^2+1\right)^2},\\
A_3&=\frac{\sqrt{\lambda } \left(n_x^2 (n_y+2) u+n_x \left(n_z-n_z u^2\right)+n_y u \left(n_y^2+2 n_y+n_z^2+1\right)\right)}{2 \pi  \left(u^2+1\right)^2}.
\end{align}
and the helicity flux density 
\bea 
\frac{dH}{du d\Omega}&=&\bm n\cdot(\bm A\times\dot{\bm A})=\frac{\lambda  (n_y+1)^2}{2 \pi ^2 \left(u^2+1\right)^3}.
\eea The helicity flux and helicity density are 
\begin{align}
    \frac{dH}{du}&=0,\\
    \frac{dH}{d\Omega}&=\frac{3(1+n_y)^2}{16\pi}.
\end{align} The helicity flux is zero is consistent with the fact that the magnetic helicity is conserved 
\be 
\mathcal H_{m}=\lambda.
\ee 
Now we turn to a more general solution mentioned in \cite{2015JPhA...48b5203A}
\be
{\bm b}(\mathbf{r},t)=\frac{\sqrt{\lambda}}{\pi }\frac{q\,{\bm H}_{1}+p\,{\bm H}_{2}}{(a^{2}+t^{2})^{3}}, \quad
{\bm e}(\mathbf{r},t)={\frac{\sqrt{a}}{\pi }}\,{\frac{q\,{\bm H}_{4}-p\,{\bm H}_{3}}{(a^{2}+t^{2})^{3}}}
\ee
where
\begin{align}
a&=\frac{x^2+y^2+z^2-t^2+1}{2} \\
p&=t(t^2-3a^2) \\
q&=a(a-3t^2) \\
{\bm H}_1 &= (-n xz + m y + s t)\, {\bm e}_x 
+ (-n yz - m x - l tz)\, {\bm e}_y 
+ \left( n \frac{-1 - z^2 + x^2 + y^2 + t^2}{2} + l t y \right)\, {\bm e}_z, \\
{\bm H}_2 &= \left( s \frac{1 + x^2 - y^2 - z^2 - t^2}{2} - m t y \right)\, {\bm e}_x 
+ (s xy - l z + m tx)\, {\bm e}_y 
+ (s xz + l y + n t)\, {\bm e}_z, \\
{\bm H}_3 &= (-m xz + n y + l t)\, {\bm e}_x 
+ (-m yz - n x - s tz)\, {\bm e}_y 
+ \left( m \frac{-1 - z^2 + x^2 + y^2 + t^2}{2} + s ty \right)\, {\bm e}_z, \\
{\bm H}_4 &= \left( l \frac{1 + x^2 - y^2 - z^2 - t^2}{2} - n ty \right)\, {\bm e}_x 
+ (l xy - sz + n tx)\, {\bm e}_y 
+ (l xz + s y + m t)\, {\bm e}_z.
\end{align}
The corresponding vector field at $\mathcal I^+$ is 
\begin{align}
    A_1(u,\Omega)&=\frac{\sqrt{\lambda } \left(\left(u^2-1\right) \left(l+n n_y\right) \left(n_y^2+n_z^2\right)-2 m u n_x n_z+n \left(u^2-1\right) n_x^2 n_y\right)}{2 \pi  \left(u^2+1\right)^2},\\
    A_2(u,\Omega)&=\frac{\sqrt{\lambda } \left(-\left(u^2-1\right) n_x \left(l n_y+n n_y^2+n n_z^2\right)-2 u n_z \left(m n_y+s\right)+\left(n-n u^2\right) n_x^3\right)}{2 \pi  \left(u^2+1\right)^2},\\
    A_3(u,\Omega)&=\frac{\sqrt{\lambda } \left(-l \left(u^2-1\right) n_x n_z+2 u n_y \left(m n_y+s\right)+2 m u n_x^2\right)}{2 \pi  \left(u^2+1\right)^2}.
\end{align} It follows that 
\bea 
\frac{dH}{du d\Omega}=\frac{\lambda}{2\pi^2(1+u^2)^3}[(n_x^2+n_y^2)m n+n_y m l+n_yns+(n_y^2+n_z^2)ls ].
\eea 
Therefore, we get a non-vanishing helicity flux 
\be 
\frac{dH}{du}=\frac{4 \lambda  (l s+m n)}{3 \pi  \left(u^2+1\right)^3}
\ee and the radiative helicity is also non-zero 
\be 
H=\frac{1}{2} \lambda  (l s+m n).
\ee The magnetic helicity of this solution is time-dependent \cite{2015JPhA...48b5203A}
\be 
\mathcal H_{m}=\frac{\lambda}{4}\left(ls+mn+(mn-ls)\frac{1-6t^2+t^4}{(1+t^2)^4}\right).
\ee Obviously, the radiative helicity is not equal to magnetic helicity 
\be 
H\not=\mathcal H_{m}.
\ee On the other hand, the optical helicity $\mathcal{H}_{\text{op}}$ is conserved and it is equal to the radiative helicity. This is consistent with the fact that both the radiative helicity and the optical helicity  correspond to the difference in the photon number between opposite helicities.
\section{Discriminant}\label{min}
In this appendix, we prove that the discriminant $\text{dis}(v,v_{\parallel})$ is always positive for any $0<v<1$ and thus the solution of $\lambda$ should be real. There are two cases: 
\begin{enumerate}
    \item When $0<v\le \frac{1}{3}$, the coefficient of $v_{\parallel}^2$ is always non-positive and then the discriminant is always positive.
    \item When $\frac{1}{3}<v<1$, the discriminant is positive for 
    \be 
v_{\parallel}<\frac{2\sqrt{2}v}{\sqrt{9v^2-1}}\label{vparallel}
    \ee and non-positive for 
    \be 
    v_{\parallel}\ge\frac{2\sqrt{2}v}{\sqrt{9v^2-1}}.\label{vparallel2}
    \ee Note that $v_{\parallel}\le v$ and thus the inequality \eqref{vparallel2} cannot be satisfied. 
\end{enumerate} 
\section{Integrals}\label{int}
In this appendix, we will investigate the following integrals on the unit sphere 
\bea 
I(\alpha;\bm v)=\int d\Omega (1-\bm n\cdot\bm v)^{-\alpha}
\eea where the norm of $\bm v$ is less than 1. By choosing the $z$ axis along $\bm v$, the integral can be easily evaluated as 
\be 
I(\alpha;\bm v)=\frac{2\pi}{(\alpha-1)v}[(1-v)^{1-\alpha}-(1+v)^{1-\alpha}],\quad \alpha\not=1,\quad |v|<1.
\ee When $\alpha=1$, the integral is the following limit 
\be 
I(1;\bm v)=\lim_{\alpha\to 1} I(\alpha,\bm v)=\frac{2\pi}{v}\log\frac{1+v}{1-v}.
\ee Note that the function $I(\alpha,\bm v)$ obeys the recurrence relations
\bea 
v^i\frac{\partial}{\partial v^i}I(\alpha;\bm v)&=&\alpha I(\alpha +1;\bm v)-\alpha I(\alpha;\bm v)
\eea which is rather similar to the ones for hypergeometric functions. Indeed, the function $I(\alpha;\bm v)$ can be written as a hypergeometric function 
\be 
I(\alpha;\bm v)=4\pi\ {}_2F_1\left(\frac{\alpha}{2},\frac{\alpha+1}{2};\frac{3}{2};v^2\right).
\ee One can extend the integral to the following tensors 
\bea
I_{i_1\cdots i_\ell}(\alpha;\bm v)=\int d\Omega n_{i_1}\cdots n_{i_\ell}(1-\bm n\cdot\bm v)^{-\alpha}.
\eea Interestingly, it is related to the scalar integral by partial derivatives with respect to the velocity 
\be 
I_{i_1\cdots i_\ell}(\alpha;\bm v)=\frac{1}{(\alpha-\ell)_\ell}\partial_{v_{i_1}}\cdots\partial_{v_{i_\ell}} I(\alpha-\ell;\bm v)
\ee
where we have defined the Pochhammer symbol 
\be 
(a)_n=a(a+1)\cdots (a+n-1)=\frac{\Gamma(a+n)}{\Gamma(a)}.
\ee 
We will also need the integral as follows 
\bea 
I(\alpha_1,\alpha_2;\bm v_1,\bm v_2)=\int d\Omega (1-\bm n\cdot\bm v_1)^{-\alpha_1}(1-\bm n\cdot\bm v_2)^{-\alpha_2},\quad |\bm v_{1,2}|<1.\label{intalpha}
\eea Several recursion relations are 
\bea 
v_1^i\frac{\partial}{\partial v_1^i} I(\alpha_1,\alpha_2;\bm v_1,\bm v_2)&=&\alpha_1 I(\alpha_1{ +1},\alpha_2;\bm v_1,\bm v_2)-\alpha_1 I(\alpha_1,\alpha_2;\bm v_1,\bm v_2),\label{rec1}\\
v_2^i\frac{\partial}{\partial v_2^i} I(\bm v_1,\bm v_2;\alpha_1,\alpha_2)&=&\alpha_2 I(\alpha_1,\alpha_2{+1};\bm v_1,\bm v_2)-\alpha_2 I(\alpha_1,\alpha_2;\bm v_1,\bm v_2),\label{rec2}\\
v_2^i\frac{\partial}{\partial v_1^i}I(\alpha_1,\alpha_2;\bm v_1,\bm v_2)&=&\alpha_1 I(\alpha_1+1,\alpha_2;\bm v_1,\bm v_2)-\alpha_1 I(\alpha_1+1,\alpha_2{ -1};\bm v_1,\bm v_2),\\
v_1^i\frac{\partial}{\partial v_2^i}I(\alpha_1,\alpha_2;\bm v_1,\bm v_2)&=&\alpha_2 I(\alpha_1,\alpha_2+1;\bm v_1,\bm v_2)-\alpha_2I(\alpha_1{ -1},\alpha_2+1;\bm v_1,\bm v_2).
\eea 
We also have the relations 
\bea 
I(0,\alpha_2;\bm v_1,\bm v_2)&=& I(\alpha_2;\bm v_2),\quad I(\alpha_1,0;\bm v_1,\bm v_2)=I(\alpha_1;\bm v_1).
\eea 
Using the method for Feynman integrals, we find 
\bea 
&{}&I(\alpha_1,\alpha_2;\bm v_1,\bm v_2)=\frac{\Gamma(\alpha_1+\alpha_2)}{\Gamma(\alpha_1)\Gamma(\alpha_2)}\int d\Omega \int_0^1 dx \frac{x^{\alpha_1-1}(1-x)^{\alpha_2-1}}{(1-x \bm n\cdot\bm v_1-(1-x)\bm n\cdot\bm v_2)^{\alpha_1+\alpha_2}}\nn\\&=&\frac{\Gamma(\alpha_1+\alpha_2)}{\Gamma(\alpha_1)\Gamma(\alpha_2)}\int_0^1 dx x^{\alpha_1-1}(1-x)^{\alpha_2-1}I(x \bm v_1+(1-x)\bm v_2;\alpha_1+\alpha_2)\nn\\&=&4\pi \frac{\Gamma(\alpha_1+\alpha_2)}{\Gamma(\alpha_1)\Gamma(\alpha_2)}\int_0^1 dx x^{\alpha_1-1}(1-x)^{\alpha_2-1}
{}_2F_1\left(\frac{\alpha_1+\alpha_2}{2},\frac{\alpha_1+\alpha_2+1}{2},\frac{3}{2};(x\bm v_1+(1-x)\bm v_2)^2\right)\nn\\\eea
For $\alpha_1=\alpha_2=1$, the integration can be found as 
\bea 
I(1,1;\bm v_1,\bm v_2)&=&4\pi\int_0^1 dx \frac{1}{1-(x \bm v_1+(1-x)\bm v_2)^2}\nn\\&=&\frac{2\pi}{\sqrt{\Delta}}\log \frac{1-\bm v_1\cdot\bm v_2+\sqrt{\Delta}}{1-\bm v_1\cdot\bm v_2-\sqrt{\Delta}}
\eea where 
\be 
\Delta=(\bm v_1-\bm v_2)^2+(\bm v_1\cdot\bm v_2)^2-\bm v_1^2\bm v_2^2.
\ee Note that 
\be 
I(1,1;\bm v,\bm v')=\int d\Omega (1-\bm n\cdot\bm v)^{-1}(1-\bm n\cdot\bm v')^{-1}
\ee is exactly the one we need to compute the radiative energy from \eqref{energyfluxtwo}. The integral \eqref{intalpha} can be extended further 
\bea 
I_{i_1\cdots i_\ell}(\alpha_1,\alpha_2;\bm v_1,\bm v_2)=\int d\Omega \frac{n_{i_1}\cdots n_{i_\ell}}{(1-\bm n\cdot\bm v_1)^{\alpha_1}(1-\bm n\cdot\bm v_2)^{\alpha_2}}
\eea which is related to \eqref{intalpha} by the identity
\bea 
I_{i_1\cdots i_\ell}(\alpha_1,\alpha_2;\bm v_1,\bm v_2)=\frac{1}{(\alpha_1-p)_p (\alpha_2-\ell+p)_{\ell-p}} \frac{\partial^p}{\partial v_1^{i_1}\cdots\partial v_1^{i_p}}\frac{\partial^{\ell-p}}{\partial v_2^{i_{p+1}}\cdots\partial v_2^{i_{\ell}}}I(\alpha_1-p,\alpha_2-\ell+p;\bm v_1,\bm v_2)\nn\\
\eea where $p=0,1,\cdots,\ell$.



\bibliography{biblio}
\end{document}